 \documentclass[10pt]{aastex63}		
 \usepackage{natbib}
\def\a{\"a}
\def\o{\"o}
\def\u{\"u}
\def\A{\"A}
\def\O{\"O}
\def\U{\"U}

\def\ang{\AA}
\def\arcsec{\hbox{$^{\prime\prime}$}}

\def\gapprox{\lower.4ex\hbox{$\;\buildrel >\over{\scriptstyle\sim}\;$}}
\def\lapprox{\lower.4ex\hbox{$\;\buildrel <\over{\scriptstyle\sim}\;$}}

\shortauthors{Aschwanden}
\shorttitle{How the SDO Revolutionized our Understanding of the Sun}

\begin{document}
\renewcommand{\topfraction}{0.95}
\renewcommand{\bottomfraction}{0.95}
\renewcommand{\textfraction}{0.05}
\renewcommand{\floatpagefraction}{0.95}
\renewcommand{\dbltopfraction}{0.95}
\renewcommand{\dblfloatpagefraction}{0.95}


\title{How the Solar Dynamic Observatory
	Revolutionized our Physical Understanding of the Sun}
 
\author{Markus J. Aschwanden}
\affil{Lockheed Martin, Solar and Astrophysics Laboratory (LMSAL),
       Advanced Technology Center (ATC),
       A021S, Bldg.252, 3251 Hanover St.,
       Palo Alto, CA 94304, USA;
       e-mail: markus.josef.aschwanden@gmail.com}

\begin{abstract}
We review solar studies using AIA, HMI, and EVE data 
from the SDO spacecraft that revolutionized our physical 
understanding of the Sun. The relevant SDO studies
cover the entire 15-year lifetime of SDO, from 2010 May 1
to 2025 May 1. The discussed phenomena and their physical
interpretations include (in chronological order):  

(1) MHD Waves and Oscillations (AIA, HMI);

(2) Propagating MHD Waves (AIA);

(3) Coronal Loop Cross-Sectional Temperatures (AIA);

(4) Size Distributions of Solar Flare Parameters (AIA);
 		
(5) Spatio-Temporal Evolution and Diffusion (AIA); 

(6) The Rosner-Tucker-Vaiana (RTV) Scaling Law (AIA);

(7) The Fractal-Diffusive Self-Organized Criticality Model (AIA);

(8) Automated Temperature and Emission Measure Maps (AIA);

(9) Automated Pattern Recognition Codes (AIA);

(10) Kelvin-Helmholtz Instability in Reconnetion Outflows (AIA);

(11) Hydrodstatics of Coronal Loops (AIA); 

(12) Magnetic Energy Dissipation (HMI);

(13) Global Energetics of Solar Flares (AIA). 

\end{abstract}

\section{	INTRODUCTION 				}  

As a spin-off of the 15$^{th}$ anniversary of the launch
of the {\sl Solar Dynamics Observatory (SDO)}, a special
issue on the topics {\sl ``How SDO has revolutionized our
understanding of the Sun''} has been planned for publication
in the journal {\sl Solar Physics}. The SDO consists of
three instruments, the {\sl Atmospheric Imaging Assembly (AIA)}
(Lemen et al.~2012),
the {\sl Helioseismic Magnetic Imager (HMI)} 
(Scherrer et al.~2012), and the
{\sl Extreme Ultraviolet Variability Experiment (EVE)}
(Woods et al.~2012).
The SDO is the first space-based mission of NASA's
{\sl Living With a Star (LWS)} program. The most useful
capability of the SDO mission is the ``imaging of all Sun
all the time''. SDO was launched on 2010 February 11,
and was lifted into a circular geosynchronous orbit
inclined by $28^\circ$ about the longitude of the
SDO-dedicated ground station in New Mexico. Science
operations started on 2010 May 1. 

The AIA onboard SDO consists of four telescopes
that employ normal-incidence, multilayer-coated optics to provide
narrow-band imaging of seven extreme ultra-violet (EUV) band passes
centered mostly on iron lines: Fe XVIII (94 \ang ), Fe VIII and XXI
(131 \ang ), Fe IX (171 \ang ), Fe XII and XXIV (193 \ang ),
Fe XIV (211 \ang ), He II (304 \ang ), and Fe XVI (335 \ang ).
The HMI onboard SDO is a 
vector magnetograph that is designed to measure the Doppler shift,
intensity, and vector magnetic field at the solar photosphere using
the 6173 \ang\ Fe I absorption line.
The EVE instrument onboard SDO measures
the solar EUV irradiance from 0.1 to 105 nm (1-1050 \ang ) with
unprecedented spectral resolution (0.1 nm), temporal cadence
(10 s), and accuracy (20\%). 

In order to answer the scientific question ``how SDO has
revolutionized our understanding of the Sun'' we select
13 physical topics that are based on AIA and HMI data 
and were published during the year span of 2010-2025. 
The physical topics include coronal loops (MHD waves 
and oscillations, loop cross-sections, hydrostatics
of coronal loops), solar flare observations (spatio-temporal 
evolution, size distributions, magnetic energy dissipation,
global energetics of flares), solar flare models
(RTV scaling law, Kelvin-Helmholtz instability,
magnetic reconnection models), 
and {\sl artificial intelligence (AI)} (automated
measurements of the
emission measure, temperature, and loop geometry). 
We are aware that this selection of physical topics is 
biased towards measurements of quantitative physical 
parameters, rather than qualitative descriptions 
of observations, but should ultimately reveal the
most likely theoretical model by elimination of 
inadequate models. 

Textbooks on the physics of the solar corona can be found
in Aschwanden (2004; 2019a), which also contain 
comprehensive references to relevant reviews. 
The contents of this study includes a discussion section
of the 13 selected physical topics in sequential order,
followed by the Conclusions.

\section{	DISCUSSIONS  				}  

\subsection{MHD Waves and Oscillations}

Solar coronal loop oscillations have been observed 
with the {\sl Atmospheric Imaging Assembly (AIA)} 
and the {\sl Helioseismic and Magnetic Imager (HMI)} onboard 
the {\sl Solar Dynamic Observatory (SDO)} for the first time
during the the event of 2010 October 16, 19:05-19:35 UT
(Aschwanden and Schrijver 2011).
The magnitude of this event is a M2.9 GOES-class flare.
The heliographic coordinates of the oscillation event  
could be triangulated from the two {\sl extreme-ultraviolet 
imagers (EUVI)} of the simultaneously observing STEREO spaceraft.
This way the 3-D coordinates [x(t),y(t),z(t)] of the 
oscillating loops could be retrieved, which provides a
better accuracy than the generally available 2-D coordinates
[x(t),y(t)]. The oscillation event reveals a number of 
unexpected features (Fig.~1), such as: (i) excitation of kink-mode 
oscillations in vertical polarization (in the loop plane);
(ii) coupled cross-sections and density oscillations, with 
identical periods; (iii) no detectable kink amplitude damping 
over the observed duration of four kink-mode periods ($P=6.3$ min); 
(iv) multi-loop oscillations with slightly ($\approx 10\%$) 
different periods; and (v) a relatively cool loop
temperature of $T\approx 0.5$ MK. The electron density ratio 
external and internal to the oscillating loop can be used to
derive the kink-mode period, i.e., $n_e/n_i=(v_A/v_{Ae})^2=0.08\pm0.01$,
from the ratio of Alfv\'enic speeds deduced from the flare trigger delay.
The coupling of the kink mode and cross-sectional oscillations can be 
explained as a consequence of the loop length variation in the vertical 
polarization mode. Modeling the magnetic field in the oscillating loop 
using HMI/SDO magnetogram data and a potential field model, one finds 
agreement with the seismological value of the magnetic field, 
$B_{kink}=4.0\pm0.7$ G, within a factor of two. The data anaysis
of this event revolutionized our physical understanding of
coronal seismology by means of full 3-D modeling, where the magnetic
field ${\bf B}(x,y,z)$ is independently constrained by the
Alfv\'enic speed, $v_A=2.18 \times 10^{11} B (\mu n_i)^{-1/2}$.

The new field of {\sl coronal seismology} studies 
{\sl magneto-hydrodynamic (MHD)} waves in the solar
corona, in contrast to the established field of
{\sl helioseismology}, which deals with waves
in the interior of the Sun. Coronal waves can be
subdivided into standing MHD oscillations and into 
propagating MHD waves (Table 1). Furthermore, each
MHD wave type can be subdivided into 
fast and slow modes in their wave dispersion solution,
where the slow mode propagates with acoustic speed
$(c_s)$, and the fast mode propagates with Alfv\'enic
speed ($v_A$). The combination of transverse and
longitudinal waves leads to torsional oscillations
of helically twisted magnetic field lines 
(Aschwanden 2012a; Aschwanden and Wang~2020). 
The theoretical foundation of coronal seismology
deals with standing waves in high-temperature plasmas,
the wave dispersion relation, and wave solutions in
homogeneous plasmas and in wave ducts, which
leads to the detection of all theoretically
predicted MHD modes (Roberts et al.~1984) and 
has been well established by the work of Roberts (2019),
but only the full 3-D stereoscopic observations 
(e.g., Aschwanden and Schrijver 2011) have closed 
the gap between theory and observations, which 
can be considered as a revolutionary step in our
physical understanding of the Sun. 

\subsection{Propagating MHD Waves (AIA)}

Quasi-periodic, propagating fast-mode magnetoacoustic waves
in the corona were difficult to observe in the past due to
relatively low instrument cadences, but are important for
the understanding of the flare dynamics. The AIA/SDO has high
cadences up to 10 s, short exposures of 0.1 -- 2 s, and
a $41'\times41'$ field-of-view at $0.5\arcsec$ resolution.
Among the fastest MHD wave detections,
direct imaging by SDO/AIA of quasi-periodic fast propagating
waves with velocities of $\approx 2200\pm130$ km s$^{-1}$
(Fig.~2) were reported by Liu et al.~(2011).
Quasi-periodic fast-mode wave trains within a 
{\sl global EUV wave} and sequential transverse oscillations
were detected by AIA/SDO (Liu et al.~2012). 
The initial velocities up to 1400 km s$^{-1}$ decelerated 
to $\approx 650$ km s$^{-1}$, as a result the CME evolution
(Liu et al.~2012). 
MHD simulations of super-fast magnetosonic
waves observed by AIA/SDO in active region funnels
demonstrated that the excited fast magnetosonic waves
have similar amplitudes, wavelengths, and propagation
speeds as the observed wave trains (Ofman et al.~2011).
An overview on coronal seismology with propagating MHD waves
is given in Table 1 (Aschwanden 2012a).

\subsection{Coronal Loop Cross-Sectional Temperatures}	

The dominant geometrical structure of the solar corona 
consists of {\sl loops}, which can be quantified
in terms of open and/or closed field lines of a dipolar
magnetic field model. While the loop lenghts can readily
be determined by reconstruction of their 3-D geometry
$[x(s), y(s), z(s)]$ (or by stereoscopic reconstruction), 
the width of coronal loops is harder to disentangle,
since it depends sensitively on the instrumental
(spatial) resolution. Theoretically one expects
multi-thermal loops for unresolved loop strands,
and vice versa isothermal loops for spatially resolved
loop widths. The multi-thermality is
traditionally determined from the analysis of
the emission measure $EM(s)$ and electron 
temperature $T(s)$ along the loop axis coordinate $s$.
AIA/SDO is most suitable for this task, if the loop
unrelated background flux is properly subtracted out.
Such an analysis on the cross-sectional temperature
structure yielded near-isothermal loop cross-sections,
which is not consistent with multi-thermal nanoflaring,
but can be explained by flare-like heating mechanisms
that drive chromospheric evaporation and upflows of
heated plasma coherently over cross-sections of
$w \approx 2-4$ Mm (Aschwanden and Boerner 2011).
While some studies were consistent with isothermal
loop strands (Aschwanden and Nightingale 2005),
other studies indicated mostly multi-thermal
loop strands (Mulu-Moore et al.~2011). 
These apparently contradictory results, however,
can easily be reconciled by the degree of
applied multi-thermal background subtraction.
The key to the solution is, however, testing with
higher spatial resolution. Such an instrument with
the highest ever achieved spatial resolution of
$w \approx 0.2\arcsec$ (140 km) is the EUV imager Hi-C,
which indeed revealed mostly spatially resolved
magnetic braids (Cirtain et al.~2013). 
Automated loop cross-section detections with
AIA/SDO yields distributions of loop widths
that are consistent with Hi-C, with a peak
at $\approx 550$ km (Aschwanden and Peter 2017).
A synopsis of reported loop widths is given
in Fig.~(3).

\subsection{Size Distributions of Flare Parameters}

Solar flares span about 9 orders of magnitude
from nanoflare events of $E_{min} \approx 10^{24}$ [erg]
to the largest giant flares with energies of 
$E_{max} \approx 10^{33}$
[erg], revealing power law distribution functions
$N(E) \propto E^{-\alpha}$ with some characteristic
slope of $\alpha$. Given this highly nonlinear
dynamic range, it thus makes no sense to talk about
mean values or standard deviations of flare energies,
as it is commonly used for statistics of linear processes. 
It is more useful to characterize power law
functions and their slopes instead. There are a number of
solar flare parameters that display power law behavior,
such as the geometric parameters of length scales $L$, 
areas $A$, and volumes $V$, or physical parameters of
fluxes $F$, fluences $F$, energies $E$, electron
densities $n_e$, electron temperatures $T_e$, or
emission measures $EM$, etc. One of the first power law
distributions of solar flare parameters observed with
AIA/SDO (Aschwanden and Shimizu 2013),
is shown in Fig.~(4) and in Table 2.
Size dstributions of the corresponding geometric parameters 
have been inferred from AIA/SDO data in Aschwanden et al.~(2013a).
In the following we will see that the statistical size
distributions discussed here play a fundamental role 
in the solar flare energy budget and identifiation of 
coronal heating processes.
	
Extending to all (seven) AIA/SDO coronal wavelengths,
94, 131, 171, 193, 211, 304, and 335 \ang\ 
one finds nearly identical power laws for each wavelength,
except for the 171 and 193 \ang\ wavelengths, which are
all affected by EUV dimming caused by CMEs
(Aschwanden et al.~2013a).
The range of spatial flare sizes covers $2L \approx
10-200$ Mm, which is also the size range
of active regions on the Sun.

\subsection{Spatio-Temporal Evolution and Diffusion} 

The spatio-temporal evolution of a solar flare 
typically consists of an initial exponentially-growing
rise phase, followed by a subsequent diffusive decay phase,
in the observed flux $F(t)$. The time-integrated
flux (or fluence), which is a monotonically
increasing function, can be quantified
by fitting a radial expansion model in terms
of a generalized diffusion function,
$r(t)\propto \kappa (t-t_1)^{\beta/2}$, which
includes the logistic growth limit $(\beta=0)$,
sub-diffusion $(\beta=0-1)$, classical diffusion
$(\beta=1)$, super-diffusion $(\beta=1-2)$,
and the linear expansion limit $(\beta=2)$
(Fig.~5).
From the statistics of 155 M- and X-class
AIA-observed flares, a mean value of
$(\beta=0.53\pm0.27)$ is found, which is in
the sub-diffusive regime (Aschwanden et al.~2012b).
This result is interpreted in terms of
anisotropic chain reactions of intermittent
magnetic reconnection episodes, in a low
plasma-$\beta$ corona (where the thermal
pressure is smaller than the magnetic pressure).
A mean propagation speed of $v=15\pm12$ km s$^{-1}$
is observed, with maximum speeds of $v_{max}
=80\pm85$ km s$^{-1}$, which is substantially
slower than the magnetio-acoustic speed
expected for thermal diffusion of flare plasmas
(Aschwanden et al.~2012b). 

Alternatively, the 
diffusive character of the spatio-temporal 
propagation in solar flares has also been 
modeled in terms of cellular automaton schemes 
(Lu and Hamilton 1991).

\subsection{The Rosener-Tucker-Vaiana (RTV) Scaling Law}

The {\sl Rosner-Tucker-Vaiana (RTV)} scaling law
(Rosner et al.~1978) describes a hydrostatic equilibrium
solution of a coronal loop that is steadily and spatially
uniformly heated, has a constant pressure, and is in
equilibrium between the volumetric heating rate and the
losses by radiation and thermal conduction, yielding a
scaling law between the loop maximum temperature $T$ at the
apex, the (constant) pressure, and the loop half length $L$,
i.e., $T^2 \propto n_e L$, 
as well as a scaling law for the (constant) volumetric
heating rate $H$, 
i.e., $H \propto T^{7/2} L^{-2}$, (Aschwanden and Shimizu 2013).
The spatio-temporal evolution of the electron temperature
$(T_e)$ and electron density $(n_e)$ is shown in Fig.~(6),
as calculated from an analytical solution of the RTV model.
The correlations predicted by the RTV scaling law are
compelling, as shown in Fig.~(6e) of Aschwanden and Shimizu (2013),
so that the thermal energies $E_{th}$ can be relyably predicted
within a factor of $\approx 2$ within an energy range of 
$E_{th}=10^{29}-10^{32}$ [erg].

\subsection{The Fractal-Diffusive Self-Organized Criticality Model}

The concept of {\sl self-organized criticality (SOC)} was
originally proposed by Bak, Tang, and Wiesenfeld (1987) in a
seminal article in Physical Review Letters as an explanation
of 1/f noise and has been cited in over 10,000 publications
since. The archetype (or SOC prototype) is the sandpile, which
can be explored by numerical simulations, as well as
by physical experiments. Applications can be found in 
landscape formations, earthquakes, solar flares, 
stellar flares, black holes, 
mass extinctions, brains, river networks, traffic jams, etc. 
(Aschwanden 2011, 2022a, 2025; Aschwanden and Guedel 2021;
Bak 1996).
It has been proclaimed that power laws are the hallmarks of SOC,
which motivates the study of size distribution functions
(or occurrence frequency distributions) to identify SOC
phenomena. It was further stated that {\sl ``fractals in
nature originate from self-organized criticality dynamical
processes''} (Bak and Chen 1989), which points to
fractal geometries. To clarify the nomenclature, terms
like avalanches, catastrophes, disasters, explosive
events, flare events, cataclysimic events, or epidemics, are
interchangeably used in the SOC literature.  What is
common to all these SOC phenomena is an initial
exponential-growth evolution in the rise phase,
and is followed by a subsequent fractal-diffusive
decay phase. The microscopic spatio-temporal
evolution of a SOC avalanche can be studied 
by computer simulations of iterative next-neighbor
interactions in a discretized lattice grid, such
as in the original {\sl Bak-Tang-Wiesenfeld (BTW)} model,
or alternatively by fitting macroscopic physical
scaling laws to observed size distributions,
such as in the novel {\sl fractal-diffusive
self-organized criticality (FD-SOC)} model
(Aschwanden 2014; 2015a; 2019b; 2021; 
Aschwanden and Shimizu 2013;
Aschwanden and Gogus 2025;
Aschwanden et al.~2013, 2014; 2016.

The FD-SOC model is based on four fundamental assumptions: 
(i) fractal dimensions of spatial length scales, areas, and volumes; 
(ii) scale-freeness of spatial scales;
(iii) flux-volume proportionality for incoherent fluxes; and
(iv) classical diffusive transport.

Each fractal domain has a maximum fractal dimension of $D_d=d$,
a minimum value of $D_d=(d-1)$, and a mean value of $D_V=d-1/2$,
\begin{equation}
        D_V={(D_{\rm V,max} + D_{\rm V,min}) \over 2} = d-{1 \over 2}  \ .
\end{equation}
For most applications in the observed 3-D world, the dimensional
domain $d=3$ is appropriate, which implies a fractal dimension of
$D_V=2.5$. The fractal volume $V$ is then defined by the standard
(Hausdorff) fractal dimension $D_V$ in 3-D and the length scale
$L$ (Mandelbrot 1977),
\begin{equation}
        V(L) \propto L^{D_V} \ .
\end{equation}
The flux is assumed to be proportional to the avalanche volume
for the case of incoherent growth ($\gamma=1$),
but can be generalized for coherent growth ($\gamma \gapprox 2$),
\begin{equation}
        F \propto V^\gamma = \left( L^{D_V} \right)^\gamma \ .
\end{equation}
The spatio-temporal evolution is approximated with the assumption
of (classical) diffusive transport,
\begin{equation}
        L \propto T^{\beta/2} = T^{1/2} \ ,
\end{equation}
with the diffusion coefficient $\beta=1$.
The statistics of SOC avalanches is quantified in terms of size
distributions (or occurrence frequency
distributions) that obey the scale-free probability distribution
function (Aschwanden 2014, 2015a, 2022a), expressed with the power
law function, akas Zipf's law (Newman 2005),
\begin{equation}
        N(L)\ dL \propto L^{-d} dL \ .
\end{equation}
From the scale-free relationship, the power law slopes $\alpha_s$
of other SOC size parameters $s=[A,V,F,E,T]$ can be derived,
such as for the area $A$, the volume $V$, the flux $F$,
the fluence or energy $E$, and the duration $T$. The resulting
power law slopes $\alpha_s$ can then be obtained mathematically
by the method of variable substitution $s(L)$, by inserting the
inverse function $L(s)$ and its derivative $|dL/ds|$,
\begin{equation}
        N(s) ds = N[L(s)] \left| {dL \over ds} \right| dL
        = \ s^{-\alpha_s} ds \ ,
\end{equation}
such as for the flux $s=F$,
\begin{equation}
        \alpha_F = 1 + {(d-1) \over D_V \gamma} = {9 \over 5} = 1.80 \ ,
\end{equation}
for the fluence of energy $s=E$,
\begin{equation}
        \alpha_E = 1 + {(d-1) \over d \gamma} = {5 \over 3} = 1.67 \ .
\end{equation}
and for the time duration $s=T$,
\begin{equation}
        \alpha_T = 1 + (d-1)\beta/2 = 2 \ .
\end{equation}
We call this model the standard FD-SOC model,
defined by [$d=3, \gamma=1, \beta=1$],
while the generalized FD-SOC model allows for variable coefficients
[$d, \gamma, \beta$].
For solar and stellar flare energies, the prediction of the 
FD-SOC model is the power law slope $\alpha_E=1.67$ (Fig.~7).
While the synthesized size distributions
shown in Fig.~(7) are assembled by alternative instruments, 
in order to illustrate the large energy range, similar results 
could be furnished by the AIA/SDO instrument (see Fig.~4f;
$\alpha_{Eth}=1.66$), where the theory agrees well with the
observations.

\subsection{Automated Temperature and Emission Measure Maps}

AIA observes the Sun in six coronal temperature filters and the images
in each wavelength contain the EUV intensities $F_{\lambda}(x,y)$
that emphasize those coronal features with temperatures near
the peak sensitivity $T_{\lambda}$ of the filter passband.
A synthesized view of the six different filters can be conveyed by
the differential emission measure [DEM] distribution 
$\mathrm{d}EM(T,x,y)/\mathrm{d}T$,
which can be reconstructed from the six filter fluxes [$F_{\lambda}(x,y)$]
(in each pixel), 
\begin{equation}
        F_{\lambda}(x,y) = \int {\mathrm{d}EM(T,x,y) \over \mathrm{d}T}
        R_{\lambda}(T) \ \mathrm{d}T \ ,
\end{equation}
where the flux [$F_{\lambda}(x,y)$] is defined in units of data numbers 
per second [DN s$^{-1}$ pixel$^{-1}$],
$R_{\lambda}(T)$ is the filter-response function (in units
of DN cm$^5$ s$^{-1}$ pixel$^{-1}$), and $\mathrm{d}EM/\mathrm{d}T$ 
is the emission measure [in units of cm$^{-5}$ K$^{-1}$].
One of the simplest representations of a DEM distribution is a Gaussian
function, 
\begin{equation}
        {\mathrm{d}EM(T,x,y) \over \mathrm{d}T}= EM_{p}(x,y)
                \exp{\left(- {[\log(T)-\log(T_{p}(x,y))]^2
                \over 2 \sigma_T^2(x,y) } \right)} \ ,
\end{equation}
that can be characterized by the three parameters [$EM_{p}$, $T\_{p}$, 
and $\sigma_{p}$] for each pixel position [$(x,y)$]. A single Gaussian 
function was found to fit background-subtracted coronal loops in 
66\% of the cases
with a goodness-of-fit of $\chi^2 \le 2$ (Aschwanden and Boerner, 2011).
Even if a DEM is generally a more complicated function, it is useful to
characterize
it with the DEM peak emission-measure value [$EM_{p}$] and DEM peak
temperature [$T_{p}$], which can be achieved by a Gaussian fit. With this
method we can produce an emission-measure map [$EM_{p}(x,y)$] and a
temperature map [$T_{p}(x,y)$] (Fig.~8), but we have to keep in mind that the
observed plasma is not necessarily isothermal for any position [$(x,y)$], 
but rather characterized by an {\sl emission measure-weighted 
temperature}.

A numerical code designed for automated analysis of
SDO/AIA image datasets in the six coronal filters has been
developed (Aschwanden et al.~2013b), which accomplishes
\textit{(1)} coalignment test between different wavelengths with 
measurements of the altitude of the EUV-absorbing chromosphere,
\textit{(2)} self-calibration by empirical correction of
instru\-men\-tal-response function,
\textit{(3)} automated generation of differential emission measure [DEM]
distributions with
peak temperature maps [$T_{p}(x,y)$] and
emis\-sion-measure maps [$EM_{p}(x,y)$] (Fig.~8) 
of the full-Sun or active-region areas,
\textit{(4)} composite DEM distributions [$\mathrm{d}EM(T)/\mathrm{d}T$] of active regions or subareas;
\textit{(5)} automated detection of coronal loops, and
\textit{(6)} automated background subtraction and thermal analysis of coronal loops,
which yields statistics of loop temperatures [$T_e$], temperature widths
[$\sigma_T$], emission measures [$EM$], electron densities [$n_e$], and
loop widths [$w$]. The combination of these numerical codes allows for
automated and objective processing of a large number of coronal loops.
As an example, we present the results of an application to the active
region NOAA 11158, observed on 15 February 2011, shortly before it 
produced the largest (X2.2) flare during the current solar cycle 
(Aschwanden et al.~2014b).
A total of 570 loop segments were detected at temperatures in the 
entire range of $\log(T_e)$ = 5.7\,--\,7.0\,K and corroborate previous 
TRACE and AIA results
on their near-isothermality and the validity of the Rosner--Tucker--Vaiana
(RTV) law at soft X-ray temperatures ($T \gapprox 2$ MK).
The temperature compatibility between AIA/SDO, GOES, and
RHESSI has been tested in detail with benchmark tests 
(Ryan et al.~2014; Aschwanden et al.~2015b). 
A similar code that produces automated temperature
and emission measure maps based on positive definite DEM solutions 
has also been developed independently (Cheung et al.~2015).

While the AIA 195 \ang\ channel has a dual sensitivity to 
temperatures of $T_{AIA,195} \approx 1.5$ MK as well as 15 MK,
the STEREO/EUVI instrument has also a similar dual sensitivity 
of $T_{EUVI,195} \approx 1.5$ and $\gapprox 10$ MK, and thus the
soft X-ray fluxes of major flares far behind the limb can be
estimated from STEREO/EUVI images (Nitta et al.~2013).

\subsection{Automated Pattern Recognition Codes}

{\sl Artificial Intelligence (AI) is the capability of 
computational systems to perform tasks typically 
associated with human intelligence, such as learning, 
reasoning, problem-solving, perception, and decision-making. 
It is a field of research in computer science that develops 
and studies methods and software that enable machines to 
perceive their environment and use learning and intelligence 
to take actions that maximize their chances of achieving 
defined goals. High-profile applications of AI include 
advanced web search engines; recommendation systems; 
virtual assistants; autonomous vehicles; generative 
and creative tools; and superhuman play and analysis 
in strategy games}, accoding to Wikipedia.

Since solar coronal loops are the most conspicuous structures
in solar imaging data, a computer code with the capability of
automated detection is most desirable. Such an automated
pattern recognition that detects and extracts one-dimensional
curvi-linear features from two-dimensional digital imaging data,
called {\sl Oriented Coronal CUrved Loop Tracing (OCCULT)} code,
has been applied to TRACE, AIA/SDO, and biological data
(Aschwanden, DePontieu, and Katrukha 2013b).
Image segmentation is an image processing method that subdivides
an image into its constituent regions or objects, which can have
the one-dimensional geometry of curvi-linear (1D) segments, or the
two-dimensional (2D) geometry of (fractal) areas.
Common techniques include point,
line, and edge detection, edge linking and boundary detection, Hough
transform, thresholding, region-based segmentation, morphological
watersheds, etc. (e.g., Gonzales et al.~2008). Since there exists no
omni-potent automated pattern recognition code that works for all
types of images equally well, we have to customize suitable
algorithms for each data type individually by taking advantage
of the particular geometry of the features of interest, using
{\sl a priori} information from the data. Here we optimize
an automated pattern recognition code to extract magnetized
loops from images of the solar corona with the aim of optimum
completeness and fidelity. The particular geometric property of
the extracted features is the relatively large curvature radius
of coronal magnetic field lines, which generally do not have sharp
kinks and corners, but exhibit continuity in the variation of
the local curvature radius along their length. Using a related
strategy of curvature constraints, coronal loops were extracted also
with the directional 2D Morlet wavelet transform (Biskri et al.~2010).
In addition, solar coronal loops, as well as biological microtubule
filaments, have a relatively small cross-section compared with their
length, so that they can be treated as curvi-linear 1D objects in a
tracing method. Tracing of 1D structures with large curvature
radii simplifies an automated algorithm enormously, compared with
segmentation of 2D regions with arbitrary geometry and possibly
fractal fine structure (McAteer et al.~2005). Fig.(9) illustrates
the results of automatically traced coronal loops from an AIA/SDO
image.

\subsection{Kelvin-Helmholtz Instability in Reconnetion Outflows}

Flows and instabilities play a major role in the dynamics of
magnetized plasmas, including the solar corona, magnetospheric
and heliospheric boundaries, generation of turbulence, enhanced
aerodynamic drag, cometary tails, and astrophysical jets.
The Kelvin-Helmholtz instability (KHI) can occur in a single
continuous fluid (if there is velocity shear) or at the
interface between two fluids (if they have different velocities).
In solar physics, studies on the KHI have increased dramatically,
applied to a number of phenomena with velocity shear, such as
surges, jets, plumes, and spicules in the photosphere and
chromosphere, but also in the corona, coronal streamers,
and the solar wind. A major motivation to study the KHI
is the evolution from laminar to trubulent flows, which
affects the efficiency of coronal heating.
A particularly well-observed case with data from AIA/SDO
and clear signs of turbulent vortices in reconnection outflows 
in the lower corona was reported by Fullon et al.~(2011, 2013).
Fig.~(10) shows a train of about four vortices evolving at the
boundary between the flare ejecta and the outer ambient
solar wind.

\subsection{Hydrodstatics of Coronal Loops}

Coronal loops are curvi-linear structures aligned with the
magnetic field. The cross-section of a loop is essentially
defined by the spatial extent of the heating source
because the heated plasma distributes along the coronal
magnetic field lines without cross-field diffusion, since
the thermal pressure is much less than the magnetic
pressure in the solar corona. The solar corona consists of
many thermally isolated loops, where each one has its
own gravitational stratification, depending on its plasma
temperature. A useful quantity is the hydrostatic pressure
scale height $\lambda_p$, which depends only on the electron
temperature $T_e$,
\begin{equation}
	\lambda_p(T_e) = {2 k_B T_e \over \mu m_H G_{\odot}}
	\approx 47,000 \left({T_e \over 1\ {\rm MK}}\right) \quad [{\rm km}] \ ,
\end{equation}
where $m_H$ denotes the hydrogen mass.
Observing the solar corona in soft X-rays or EUV,
which are both optically thin emission, the line-of-sight
integrated brightness intercepts many different scale
heights, leading to a hydrostatic weighting bias toward
systematically hotter temperatures in larger altitudes above
the limb. The observed height dependence of the density
needs to be modeled with a statistical ensemble of 
multi-hydrostatic loops. Measuring a density scale height of a
loop requires careful consideration of projection effects,
loop plane inclination angles, cross-sectional variations,
line-of-sight integration, and the instrumental response
function. Hydrostatic solutions have been computed from
the energy balance between the heating rate, the radiative
energy loss, and the conductive loss. The major unknown
quantity is the spatial heating function, but analysis of
loops in high-resolution images indicate that the heating
function is concentrated near the footpoints, say at altitudes
of $h \le 20,000$ km. Of course, a large number of coronal
loops are found to be not in hydrostatic equilibrium, while
nearly hydrostatic loops have been found preferentially in
the quiet corona and in older dipolar active regions. An
example of an active region (recorded with the {\sl Transition
Region and Coronal Explorer (TRACE)} about 10 h after a
flare is shown in Fig.~(11), which clearly shows
super-hydrostatic loops where the coronal plasma is distributed
over up to four times larger heights than expected in
hydrostatic equilibrium (Fig.~11, bottom). Similar
super-hydrostatic scale heights were observed with AIA/SDO.
Aschwanden and Nitta (2000).

\subsection{Magnetic Field and Energy Dissipation}

A number of new magnetic field models based on HMI/SDO and 
AIA/SDO data have been developed over the last 15 years that 
revolutionized our physical understanding of the Sun. 
One new code is the {\sl Vertical Current Approximation
Nonlinear Force-Free Field code (VCA-NLFFF)}, which fits
3-D magnetic field lines to 2-D loop tracings (Aschwanden 2016a).
Nonlinear force-free fields fulfill the coupled equation
system of divergence-freeness,
\begin{equation}
	{\bf B} \cdot \nabla \alpha = 0 \ ,
\end{equation}
and force-freeness,
\begin{equation}
	(\nabla \times {\bf B}) = \alpha {\bf B} \ ,
\end{equation}
where the $\alpha$-parameter is invariant along
a magnetic field line of ${\bf B}$.
Differences between photopsheric extrapolation methods and
coronal forward-fitting methods provide tools to measure 
the free energy $E_{free}$ directly, which is defined by the 
difference between the nonlinear force-free $E_{N_\perp}$ and 
the potential energy $E_{P\perp}$, i.e., 
(Aschwanden et al.~2014a, 2016b). 
\begin{equation}
	E_{\perp}^{free}=E_{N,\perp}-E_P =
	{1 \over 8\pi}
	\left({\int B^2_N({\bf x}) \ dV
        - \int B^2_P({\bf x}) \ dV }\right) \ .
\end{equation} 
Although good agreement is found between the two types of 
codes for the total nonpotential and the potential energy, a 
factor of up to 4 discrepancy is found in the free energy,
and up to a factor of 10 discrepancy in the decrease of 
the free energy during flares.
The consistency of free energies measured from different 
EUV and UV wavelengths for the first time here, demonstrates 
that vertical electric currents (manifested in form of 
helically twisted loops) can be detected and measured 
from both chromospheric and coronal tracers (Aschwanden 2015b).
A more systematic study compared three different magnetic
field extrapolation codes with observed loop trajectories,
including a potential model, a nonlinear force-free (NLFF)
model based on photospheric vector data, and an NLFF model 
based on forward fitting magnetic sources with vertical
currents, were found to agree with observed loop
structures within a misalignment angle range of
$5^\circ-12^\circ$ (Warren et al.~2018).

Another (manual) method to map coronal-loop structures 
of an active region using cubic Bezier curves and its 
applications to misalignment angle analysis has been
developed by Gary et al.~(2014). 
Alternatively, solar coronal loops have been quantified
in terms of helical twisting numbers and braiding
linkage numbers (Aschwanden 2019c).

\subsection{Global Energetics of Solar Flares}

The energetic aspect of solar flares, the probably most
important quantity in the understanding of our 
``violent Sun'', has been investigated in a series of 
13 papers, referred to
as Paper I ... Paper XIII in the following, encompassing
a basic AIA/SDO dataset of $\approx 400$ M- and X-class flares.
A first estimate of magnetic energies is calculated
from the decrease of the free (magnetic) energy, using
best fits of the {\sl Vertical Current Approximation
Nonlinear Force-Free Field (VCA-NLFFF)} model,
yielding a ratio of $E_{free} / E_P = 0.01...,0.25$
with respect to the potential field magnetic energy
(Paper I; Aschwanden, Xu, and Jing (2014a). 
The thermal energy arises from the heating of the
flare plasma and has been calculated from standard
differential emission measure analysis and is found
to have an energy ratio of 
$E_{th} = E_P / 12.9 \approx E_P \times 0.08$,
(Paper II; Aschwanden and Shimizu 2013). 
Nonthermal energies are measured from hard X-ray emitting
electrons that precipitate into the chromosphere, and
reveal an energy ratio of $E_{nt} = 0.41\ E_{magn}$, 
(Paper III; Aschwanden et al.~2016b).
The kinetic energy and gravitational energy carried
by {\sl coronal mass ejections (CMEs)} amounts to energies
of $E_{CME} \approx 10^{29}...20^{30}$ [erg],
(Paper IV; Aschwanden 2016b).    
Energy closure is obtained for the so far discussed
energies, i.e., $(E_{nt}+E_{dir}+E_{CME})/E_{magn}
=0.87\pm0.18$, (Paper V; Aschwanden et al.~2017).
Further refinements in the calculation of CME energies
includes adiabatic expansion, distinction of eruptive
and confined CMEs, EUV dimming, CME leading edge speeds,
(Paper VI; Aschwanden 2017), 
as well as the aerodynamic drag force, 
(Paper VII; Aschwanden and Gopalswamy 2019b).
A most sensitive parameter is the nonthermal low-energy
cutoff, (Paper VIII; Aschwanden et al.~2019). 
Further refinement of magnetic energies and their
influence in the energy closure is provided in
Paper IX (Aschwanden 2019d). 
Magnetic reconnection geometry with the Petschek model
is considered to be more realistic than the
Sweet-Parker current sheet model
(Paper X, Aschwanden 2020b).
The magnitude of the soft X-ray flux as observed with GOES
can be predicted based on their correlation,
(Paper XI, Aschwanden 2020c).
Scaling laws of solar flare parameters resulted into
the Rosner-Tucker-Vaiana and the Shibata-Yokoyama models,
(Paper XII, Aschwanden 2020a).
The latest addition is the Neupert effect, which 
approximates the fluence with the time-integated flux
(Paper XIII, Aschwanden 2022b).
A pie-chart representation of the energy closure is
provided in Fig.~(12).

\section{	CONCLUSIONS 				}  

We selected 13 scientific topics in solar physics 
that revolutionized our physical understanding of the Sun
in the sense that new methods and new concepts have been 
applied to SDO data. 
The physical topics include coronal loops (MHD waves
and oscillations, loop cross-sections, hydrostatics
of coronal loops), solar flare observations (spatio-temporal
evolution, size distributions, magnetic energy dissipation,
global energetics of flares), solar flare models
(magnetic reconnection, Sweet-Parker model, 
Shibata-Yokoyama model, RTV scaling law, Kelvin-Helmholtz 
instability), and {\sl artificial intelligence (AI)} 
(automated measurements of the emission measure, 
temperature, and loop geometry). Here are some guiding
physical principles that emerged from these topics:

\begin{enumerate}
\item{\underbar{Coronal loop physics:} Coronal loops are
tracers of magnetic field lines, and thus should be modeled
in terms of 3-D trajectories of nonlinear force-free magnetic
field models in low plasma-$\beta$ regions. AIA and Hi-C data
of coronal loops reveal that coronal loops have spatially
resolved loop cross-sections at scales of $w \lapprox 100$ km 
and thus are nearly isothermal. The loop cross-sections
reveal the heating cross-sections at the footpoint of coronal
loops, which can be modeled by magnetic reconnection geometries,
such as Parker nanoflares.  MHD waves and oscillations
provide unambiguous loop tracers that move on top of
non-oscillating background along the line-of-sight.
The density $n_e(s)$ and temperature profiles $T_e(s)$ along
a coronal loop should follow the hydrostatic scale height
$\lambda_T$ if the loop is in pressure equilibrium, but 
TRACE and AIA data generally suggest non-equilibrium,
with super-hydrostatic scale heights that exceed up to
5 hydrostatic scale heights.}

\item{\underbar{Solar flare scaling laws:} Scaling laws
of solar flare parameters include relationships between
the flare loop length $L$, the loop cross-section width $w$,
flare areas $A$, (fractal) flare volumes $V$, the electron
density $n_e$, the electron temperature $T_e$, 
the peak flux $F$, the fluence $E$, the thermal energy $E_{th}$,
the magnetic energy $E_{mag}$, and more. The size (differential
or cumulative) distribution of solar flare parameters 
generally follow a power law with a slope of $\alpha_F=(9/5)=1.80$ 
for the peak flux, and $\alpha_E=(5/3)\approx 1.67$ for the
(time-integrated) energy, as predicted by the {\sl fractal-diffusive 
self-organized criticality (FD-SOC)} model.}

\item{\underbar{Solar flare models:} The Rosner-Tucker-Vaiana
model is based on the assumption that solar flare loops are
in equilibrium between the heating rate on one side, and the
conductive and radiative losses on the other side, which is
generally expressed in terms of a scaling law. Another
physical model is the Kelvin-Helmholtz instability, which
arises at velocity-shear layers with velocity differences,
but only few cases have been observed so far. There are
numerous (untested) magnetic reconnection models. The most
important finding is that no energy related to the
potential magnetic field can be converted into flare energy,
while the free energy, i.e., the difference between the 
nonlinear force-free field and the potentail field,
constitutes an upper limit for the dissipated energy
in solar flares. This principle is largely consistent with
the observed flare energies.}

\item{\underbar{Artificial Intelligence (AI):} A possible
application of AI is the automated detection of coronal
loop geometries and magnetic field models, generally
known as automated pattern recognition codes.
Optimum performance requires a detailed customization
of AI codes. Magnetic field geometries can be calculated
in 3-D, using photospheric field extrapolation models
constrained by HMI/SDO data. These 3-D coronal field 
lines need to be matched by 2-D coronal loops, which 
can be constrained by AIA/SDO data. Successful 
performance of this approach has already been 
presented (e.g., Fig.~12).}

\end{enumerate}

\acknowledgments
{\sl Acknowledgements:
We acknowledge constructive and stimulating discussions
This work was partially supported by NASA contract NNX11A099G
``Self-organized criticality in solar physics'' and NASA contract
NNG04EA00C of the SDOAIA instrument to LMSAL.}

\clearpage

\section*{      References      }
\def\ref#1{\par\noindent\hangindent1cm {#1}}

\ref{Asai, A., Shimojo, M., Isobe, H., Morimoto, T., Yokoyama, T.,
        Shibasaki, K., and Nakajima, H. 
	2001, ApJ 562, L103.}
	{\sl Periodic acceleration of electrons in the 1998 
	November 10 solar flare}
\ref{Aschwanden, M.J. 
	1987, SoPh 111, 113.}
	{\sl Theory of radio pulsations in coronal loops}
\ref{Aschwanden, M.J., Fletcher, L., Schrijver, C., and Alexander, D.
        1999, ApJ 520, 880.}
	{\sl Coronal loop oscillations observed with the
	Transition region and coronal explorer} 
\ref{Aschwanden, M.J. and Nitta, N.
 	2000, ApJ 535, L59-L62.}
 	{\sl The Effect of Hydrostatic Weighting on the Vertical 
	Temperature Structure of the Solar Corona}
\ref{Aschwanden, M.J. 
	2004, {\sl Physics of the solar corona. An Introduction}, 
	Springer PRAXIS, Berlin.}
\ref{Aschwanden, M.J. and Nightingale, R.W.
 	2005, ApJ 633, 499-517.}
 	{\sl Elementary loop structures in the solar corona 
	analyzed from TRACE triple-filter images}
\ref{Aschwanden, M.J. 
	2011, {\sl Self-Organized Criticality in Astrophysics. 
	The Statistics of Nonlinear Processes in the Universe},
        Springer-Praxis: New York, 416p.}
\ref{Aschwanden, M.J. and Boerner,P.
 	2011, ApJ 732, 81 (15 pp).}
 	{\sl Solar corona loop studies with AIA: I. Cross-sectional 
	temperature structure}
\ref{Aschwanden, M.J. and Schrijver, C.J., 
	2011, ApJ 736, 102, (20pp).}
	{\sl Coronal loop oscillations observed with 
	Atmospheric Imaging Assembly - Kink mode with cross-sectional
	and density oscillations} 
\ref{Aschwanden, M.J.
 	2012a, Conf.Series Astron.Soc.Pacific (ASP) Vol. 463, p.133-145, 
	(eds. T.R.Rimmele, M.Collados Vera, et al.)} 
 	{\sl Coronal Seismology with ATST}
\ref{Aschwanden, M.J.
 	2012b, ApJ 757, 94.}
 	{\sl The spatio-temporal evolution of solar flares observed 
	with AIA/SDO: Fractal diffusion, sub-diffusion, or logistic growth ?}
\ref{Aschwanden, M.J. and Shimizu, T.
 	2013, ApJ 776, 132.}
 	{\sl Multi-Wavelength Observations of the Spatio-Temporal 
	Evolution of Solar Flares with AIA/SDO: II. Hydrodynamic 
	Scaling Laws and Thermal Energies}
\ref{Aschwanden, M.J., Zhang, J., and Liu, K.
 	2013a, ApJ 775, 23 (22p).}
 	{\sl Multi-wavelength Observations of the Spatio-temporal 
        Evolution of Solar Flares with AIA/SDO. I. Universal Scaling 
	Laws of Space and Time Parameters}
\ref{Aschwanden, M.J., Boerner,P., Schrijver, C.J., and Malanushenko, A.
 	2013b, SoPh 283, 5-30.}
 	{\sl Automated temperature and emission measure analysis of 
	coronal loops and active regions observed with AIA/SDO}
\ref{Aschwanden, M.J. 
	2014, ApJ 782, 54.}
        {\sl A macroscopic description of self-organized systems and
        astrophysical applications} 
\ref{Aschwanden, M.J., Xu, Y., and Jing,J.
 	2014a, ApJ 797, 50.}
 	{\sl Global energetics of solar flares: I. Magnetic Energies}
\ref{Aschwanden,M.J., Sun,X.D., and Liu,Y.
 	2014b, ApJ 785, 34.}
 	{\sl The magnetic field of active region 11158 during the 
	2011 February 12-17 flares: Differences between photospheric 
	extrapolation and coronal forward-fitting methods}
\ref{Aschwanden, M.J. 
	2015a, ApJ 814, 19 (25pp).}
        {\sl Thresholded power law size distributions of instabilities
        in astrophysics} 
\ref{Aschwanden, M.J.
 	2015b, ApJ 804, L20.}
 	{\sl Magnetic energy dissipation during the 2014 Mar 29 flare}
\ref{Aschwanden, M.J., Boerner, P., Ryan, D., Caspi, A., 
	McTiernan, J.M., and Warren, H.P.
 	2015a, ApJ 802:53 (20pp)}
 	{\sl Global energetics of solar flares: II. Thermal Energies}
\ref{Aschwanden, M.J., Boerner, P., Caspi, A., McTiernan, J.M., 
	Ryan, D., and Warren, H.P.
 	2015b, SoPh 290, 2733-2763.}
 	{\sl Benchmark Test of Differential Emission Measure Codes 
	and Multi-Thermal Energies in Solar Active Regions}
\ref{Aschwanden, M.J., Crosby, N., Dimitropoulou, M., Georgoulis, M.K., et al.,
	2016, SSRv 198, 47.}
        {\sl 25 Years of Self-Organized Criticality: Solar and
        Astrophysics} 
\ref{Aschwanden, M.J.
 	2016a, ApJSS 224, 25 (32pp)}
 	{\sl The Vertical Current Approximation Nonlinear Force-Free 
	Field Code - Description, Performance Tests, and Measurements 
	of Magnetic Energies Dissipated in Solar Flares}
\ref{Aschwanden,M.J.
 	2016b, ApJ 831, 105 (34pp).}
 	{\sl Global energetics of solar flares: IV. Coronal Mass 
	Ejection Energetics}
\ref{Aschwanden,M.J., Reardon,K., and Jess, D.
 	2016a, ApJ 826, 61 (18pp)}
 	{\sl Tracing the chromospheric and coronal magnetic field 
	with AIA, IRIS, IBIS, and ROSA data}
\ref{Aschwanden, M.J., O'Flannagain, A., Caspi, A., McTiernan, J.M., 
	Holman, G., Schwartz, R.A., and Kontar, E.P.
 	2016b, ApJ 832, 27, (20pp)}
 	{\sl Global energetics of solar flares: III. Nonthermal Energies}
\ref{Aschwanden, M.J.
 	2017, ApJ 847:27 (19pp)}
 	{\sl Global energetics of solar flares: VI. Refined energetics 
	of coronal mass ejection Energetics}
\ref{Aschwanden, M.J. and Peter,H.
 	2017, ApJ 840:4 (24pp).}
 	{\sl The width distribution of solar coronal loops 
	and strands - Are we hitting rock bottom ?}
\ref{Aschwanden, M.J., Caspi, A., Cohen, C.M.S., Gordon, H.,
	Jing, J., Kretzschmar, M. et al. 
	2017, ApJ 836/1, article id.17, 17pp.}
	{\sl Coronal energetics of solar flares. V. Energy closure
	and coronal mass ejections}, 
\ref{Aschwanden, M.J.
	2019a, {\sl New Millennium Solar Physics}, Astrophysics
	and Space Science Library (ASSL) 458, Springer:Berlin}
\ref{Aschwanden, M.J.
 	2019b, ApJ 880:105 (16pp).}
 	{\sl Self-organized criticality in solar and stellar flares: 
	Are extreme events scale-free ?}
\ref{Aschwanden, M.J.
 	2019c, ApJ 874, 131 (10pp)}
 	{\sl Helical twisting number and braiding linkage number 
	of solar coronal loops}
\ref{Aschwanden, M.J.
 	2019d, ApJ 885:49 (21pp)}
 	{\sl Global energetics of solar flares: IX. Refined Magnetic 
	Modeling}
\ref{Aschwanden, M.J. and Gopalswamy, N.
 	2019, ApJ 877:149 (14pp)}
 	{\sl Global energetics of solar flares: VII. Aerdodynamic 
	drag in coronal mass ejections}
\ref{Aschwanden,M.J., Kontar, E.P., and Jeffrey, N.L.S.
 	2019, ApJ 881:1 (22pp)}
 	{\sl Global energetics of solar flares: VIII. The Low-Energy 
	Cutoff}
\ref{Aschwanden, M.J.
 	2020a, ApJ 903:23}
 	{\sl Global energetics of solar flares. XII. Energy 
	scaling laws}
\ref{Aschwanden, M.J.
 	2020b, ApJ 895:13.}
 	{\sl Global energetics of solar flares: X. Petschek 
	reconnection rate and Alfven Mach number}
\ref{Aschwanden, M.J.
 	2020c, ApJ 897:16.}
 	{\sl Global energetics of solar flares. XI. Flare 
	magnitude predictions of the GOES class}
\ref{Aschwanden, M.J. and Wang, T.J. 
	2020, ApJ 891:99 (16pp).}
	{\sl Torsional Alfvenic oscillations discovered in the 
	magnetic free energy during solar flares}
\ref{Aschwanden, M.J. 
	2021, ApJ 909, 69.}
        {\sl Finite system-size effects in self-organizing criticality 
	systems} 
\ref{Aschwanden, M.J. and Guedel, M. 
	2021, ApJ 910, id.41, 16pp.}
	{\sl Self-organized criticality in sellar flares}, 
\ref{Aschwanden, M.J. 
	2022a, ApJ 934 33}
        {\sl The fractality and size distributions of astrophysical
        self-organized criticality systems}
\ref{Aschwanden, M.J.
 	2022b, astro-ph arXiv:2112.07759.}
 	{\sl Global energetics in solar flares. XIII. The Neupert 
	effect and acceleration of coronal mass ejections}
\ref{Aschwanden, M.J. 
	2025, {\sl Power Laws in Astrophysics. Self-Organzed Criticality
`	Systems}, Cambridge University Press: Cambridge.}
\ref{Aschwanden, M.J. and Schrijver, C.J.
	2025, ApJ, (in press).}
	{\sl Self-organized criticality across thirteen orders of 
	magnitude in the solar-stellar connection}
\ref{Aschwanden, M.J. and Gogus, E. 
	2025, ApJ, 978:19 (11pp).}
	{\sl Testing the universality of self-organized criticality
	in galactic, extra-galactic, and black-hole systems}
\ref{Bak, P., Tang, C., and Wiesenfeld, K. 
	1987, Physical Review Lett. 59(27), 381.}
        {\sl Self-organized criticality: An explanation of 1/f noise}
\ref{Bak, P. and Chen, K. 
	1989, Nature 342, 780-782.}
	{\sl The physics of fractals}
\ref{Bak, P. 
	1996, {\sl How Nature Works. The Science of Self-Organized 
	Criticality}, Copernicus: New York.}
\ref{Berghmans, D. and Clette, F. 
	1999, SoPh 186, 207.}
	{\sl Active region EUV transient brightenings - First 
	Results by EIT of SOHO JOP80}
\ref{Biskri, S., Antoine, J.P., Inhester, B., and Mekideche, F.
	2010, SoPh 262, 373.} 
	{\sl Extraction of Solar Coronal Magnetic Loops with the 
	Directional 2D Morlet Wavelet Transform} 
\ref{Cheung, M.C.M., Boerner, P., Schrijver, C.J., Testa, P., 
	Chen, F., Peter, H., and Malanushenko, A.
 	2015, ApJ 807, 143.}
 	{\sl Thermal Diagnostics with the Atmospheric Imaging 
	Assembly on board the Solar Dynamics Observatory: 
	A Validated Method for Differential Emission Measure Inversions}
\ref{Cirtain,J.W., Golub, L., Winebarger, A.R. et al.
 	2013, Nature 493, 501.}
 	{\sl Energy release in the solar corona from spatially 
	resolved magnetic braids},
\ref{DeForest, C.E. and Gurman, J.B. 
	1998, ApJ, 501, L217.}
	{\sl Observation of quasi-periodic compressive waves in 
	solar polar plumes}
\ref{DeMoortel, I., Ireland, J., and Walsh, R.W.
	2000, AA 355, L23-L26.}
	{\sl Observations of oscillations in coronal loops}
\ref{DeMoortel, I., Ireland, J., and Walsh, R.W. 
        2002, SoPh 209, 61.}
	{\sl Longitudinal intensity oscillations in coronal loops 
	observed with TRACE: I. Overview of measured parameters}
\ref{Emslie, A.G., Dennis, B.R., Shih, A.Y., Chamberlin, P.C., et al.
 	2012, ApJ 759, 71.}
 	{\sl Global Energetics of Thirty-eight Large Solar Eruptive 
	Events}
\ref{Foullon, C., Verwichte, E., Nakariakov, V.M., Nykyri, K., 
	and Farrugia,C.J.
 	2011, ApJ 729, L8.}
 	{\sl Magnetic Kelvin-Helmholtz instability at the Sun}
\ref{Foullon, C., Verwichte, E., Nykyri, K., Aschwanden, M.J., 
	and Hannah,I.
 	2013, ApJ 767, 170, 18pp}
 	{\sl Kelvin-Helmholtz Instability of the CME reconnection 
	outflow layer in the low corona}
\ref{Gary, G.A., Hu, Q., and Lee, J. K.
 	2014, SoPh 289, 847-865.}
 	{\sl A Rapid, Manual Method to Map Coronal-Loop Structures 
	of an Active Region Using Cubic Bezier Curves and Its 
	Applications to Misalignment Angle Analysis}
\ref{Gonzales, R.C., Woods, R.E.. 
	2008, Digital Image Processing, Pearson Prentice Hall,
	Upper Saddle River, NJ 07458, p.689-794.}
\ref{Katsiyannis, A.C., Williams, D.R., McAteer, R.T.J., 
	Gallagher, P.T., Keenan,F.P., and Murtagh, F. 
	2003, AA 406, 709.}
	{\sl Eclipse observations of high-frequency oscillations 
	in active region coronal loops}
\ref{Kliem, B., Dammasch, I.E., Curdt, W., and Wilhelm,K.,
        2002, ApJ 568, L61.}
	{\sl Correlated hot and cool plasma dynamics in the main
 	phase of a solar flare} 
\ref{Lemen, J.R., Title, A.M., Akin, D.J. et al. 
	 2012, SoPh 275, 17-40.}
 	{\sl The Atmospheric Imaging Assembly (AIA) on the Solar 
	Dynamics Observatory (SDO)}
\ref{Liu, W., Title, A.M., Zhao, J.W., Ofman, L., Schrijver, C.J., 
	Aschwanden,M.J., DePontieu,B., and Tarbell,T.T.
 	2011, ApJ 736, L13.}
 	{\sl Direct imaging by SDO/AIA of quasi-periodic propagating 
	fast mode magnetosonic waves of 2000 km/s in the solar crona}
\ref{Liu, W., Ofman, L., Nitta, N.V., Aschwanden, M.J., Schrijver, C.J., 
	Title, A.M., and Tarbell,T.D.
 	2012, ApJ 753, 52.}
 	{\sl Quasi-periodic fast-mode wave trains within a global EUV 
	wave and sequential transverse oscillations detected by SDO/AIA}
\ref{Lu, E.T. and Hamilton, R.J.
 	1991, ApJ 380, L89-L92.}
 	{\sl Avalanches and the distribution of solar flares}
\ref{Mandelbrot, B.B. 
	1977, {\sl Fractals: form, chance, and dimension}, Translation of
        {\sl Les objects fractals}, W.H. Freeman, San Francisco.}
\ref{McAteer, R.T.J., Gallagher, P.T., and Ireland, J.
 	2005, ApJ 631, 628-635.}
	{\sl Statistics of Active Region Complexity: A Large-Scale 
	Fractal Dimension Survey}
\ref{Melnikov, V.F., Reznikova, V.E., Shibasaki, K., and 
	Nakariakov, V.M.
        2005, AA 439, 727.}
	{\sl Spatially resolved microwave pulsations of a flare loop}
\ref{Mulu-Moore, F.M., Winebarger, A.R., Warren, H.P., and Aschwanden,M.J.
 	2011, ApJ 733, 59 (7pp).}
 	{\sl Determininig the temperature structure of solar coronal 
	loops using their temporal evolution}
\ref{Nakariakov, V.M., Ofman, L., DeLuca, E., Roberts, B., and 
	Davila, J.M.
        1999, Science, 285, 862.}
	{\sl TRACE observations of damped coronal loop oscillations: 
	implications for coronal heating}
\ref{Nakariakov, V.M., Melnikov, V.F., and Reznikova, V.E. 
	2003, AA 412, L7.}
	{\sl Global sausage modes of coronal loops}
\ref{Ofman, L., Liu, W., Title, A., and Aschwanden, M.
 	2011, ApJL 740, L33 (6pp).}
 	{\sl Modeling super-fast magnetosonic waves observed by 
	SDO in active region funnels}
\ref{Newman,M.E.J. 2005,
        {\sl Power laws, Pareto distributions and Zipf's law},
        Contemporary Physics, Vol. 46, issue 5, pp.323-351.}
\ref{Nitta, N.V., Aschwanden, M.J., Boerner, P.F., Freeland, S.L., 
	Lemen, J. R., and Wuelser, J.P.
 	2013, SoPh 288, 241-254.}
 	{\sl Soft X-ray Fluxes of Major Flares Far Behind the 
	Limb as Estimated Using STEREO EUV Images} 
\ref{Robbrecht, E., Verwichte, E., Berghmans, D., Hochedez, J.F., 
	Poedts, S., and Nakariakov, V.M. 
	2001, AA 370, 591.}
	{\sl Slow magnetoacoustic waves in coronal loops: EIT 
	and TRACE}
\ref{Roberts,, B., Edwin, P.M., and Benz, A.O. 
	1984, ApJ 279, 857.}
	{\sl On coronal oscillations}
\ref{Roberts, B. 
	2019, {\sl MHD Waves in the Solar Atmosphere},
	Cambridge Universith Press, Cambridge.} 
\ref{Rosner, R., Tucker, W.H., and Vaiana, G.S.
 	1978, ApJ 220, 643-665.}
 	{\sl Dynamics of quiescent solar corona}
\ref{Ryan, D.F., O'Flannagain, A.M., Aschwanden, M.J., Gallagher,P.T.
 	2014, SoPh 289, 2547-2563.}
 	{\sl The compatibility of flare temperatures observed with 
	AIA, GOES, and RHESSI}
\ref{Scherrer, P.H., Schou, J., Bush, R.I., Kosovichev, A.G., et al.
 	2012, SoPh 275, 207-227.}
 	{\sl The Helioseismic and Magnetic Imager (HMI) Investigation 
	for the Solar Dynamics Observatory (SDO)}
\ref{Tomczyk, S., McIntosh, S.W., Keil, S.L., Judge, P.G., Schad, T.,
        Seeley, D.H., and Edmondson, J. 
	2007, Nature, {\bf 317}, 1192.}
	{\sl Alfven waves in the solar corona}
\ref{Tomczyk, S. and McIntosh, S.W. 
	2009, ApJ 697, 1384.}
	{\sl Time-distance seismology of the solar corona with CoMP}
\ref{Verwichte, E., Nakariakov, V.M., and Cooper, F.C.
        2005, AA 430, L65.}
	{\sl Transverse waves in a post-flare super-arcade}
\ref{Wang, T.J., Solanki, S.K., Curdt, W., Innes, D.E., and 
	Dammasch, I.E.,
        2003, AA, 406, 1105.}
	{\sl Hot coronal loop oscillations observed with SUMER: 
	examples and statistics}
\ref{Warren, H.P., Crump, N.A., Ugarte-Urra, I., Sun, X., 
	Aschwanden, M.J., and Wiegelmann,T.
 	2018, ApJ 860, 46, (13pp)}
 	{\sl Towards a quantitative comparison of magnetic field 
	extrapolations and observed coronal loops}
\ref{Williams, D.R., Phillips, K.J.H., Rudawy, P., Mathioudakis, M.,
        Gallagher, P.T., O'Shea, E., Keenan, F.P., Read, P., Rompolt, B.
        2001, MNRAS 326, 428.}
	{\sl High frequency oscillations in a solar active region 
	coronal loops}
\ref{Woods, T.N., Eparvier, F.G., Hock, R., Jones, A.R., et al.
 	2012, SoPh 275, 115-143.}
 	{\sl Extreme Ultraviolet Variability Experiment (EVE) on the 
	Solar Dynamics Observatory (SDO): Overview of Science 
	Objectives, Instrument Design, Data Products, and Model 
	Developments}

\clearpage

\begin{table}
\begin{center}
\footnotesize
\begin{tabular}{|l|l|l|l|} \hline
MHD wave type     & Period range        & Observations  & References of examples \\ \hline
\underbar{\sl MHD Oscillations}&        &               &               \\
Fast kink mode    & $\approx 3-5$ min   & TRACE         & Aschwanden et al.~(1999) \\
                  &                     &               & Nakariakov et al.~(1999) \\
Fast sausage mode & $\approx 1-10$ s    & Radio         & Aschwanden (1987) \\
                  &                     & Nobeyama      & Asai et al.~(2001) \\
                  &                     & Nobeyama      & Nakariakov et al.~(2003) \\
                  &                     & Nobeyama      & Melnikov et al.~(2005) \\
Slow (acoustic) mode& $\approx 10-20$ min & SOHO/SUMER  & Wang et al.~(2003) \\
                  &                     & SOHO/SUMER    & Kliem et al.~(2002) \\
\hline
\underbar{\sl Propagating MHD Waves}&   Velocity range & &              \\
Slow (acoustic) waves & $75-150$ km/s   & SoHO/EIT      & DeForest \& Gurman (1998) \\
                  & $75-200$ km/s       & SoHO/EIT      & Berghmans \& Clette (1999) \\
                  & $70-235$ km/s       & TRACE         & DeMoortel et al.~(2000, 2002) \\
                  & $65-150$ km/s       & TRACE, SoHO/EIT&Robbrecht et al.~(2001) \\
Fast (Alfv\'enic) waves & $2100$ km/s   & SECIS         & Williams et al.~(2001) \\
                  &                     &               & Katsiyannis et al.~(2003) \\
                  & $1000-4000$ km/s    & CoMP          & Tomczyk et al. (2007) \\
                  &                     &               & Tomczyk and McIntosh (2009) \\
Fast kink waves   & $100-500$ km/s      & TRACE         & Verwichte et al.~(2005) \\
\hline
\end{tabular}
\caption{MHD wave types identified in the solar corona
(Aschwanden 2012a).}
\end{center}
\end{table}

\begin{table}
\begin{center}
\caption{Power law slopes observed with AIA/SDO for 155 events
(extracted from Aschwanden et al.~2013 and Aschwanden and Shimizu 2013.}
\normalsize
\medskip
\begin{tabular}{|l|l|r|r|}
\hline
Flare parameter		& Symbol	& Power law	& FD-SOC     \\
                        &               & observation 	& prediction \\
                        &               & $\alpha_x$	& $\alpha_x$ \\
\hline
Length scale 		& $L$		& $3.2\pm0.7$	& 3.00 \\
Area			& $A$		& $2.1\pm0.3$	& 2.25 \\
Volume			& $V$		& $1.6\pm0.2$	& 1.67 \\
Time duration		& $T$		& $2.1\pm0.2$	& 2.00 \\
Emission measure	& $EM$		& $1.78\pm0.03$	& 1.80 \\
Electron density	& $n_e$		& $2.15\pm0.17$	& ...  \\
Thermal energy		& $E_{th}$	& $1.66\pm0.13$	& 1.67 \\
\hline
\end{tabular}
\end{center}
\end{table}


\begin{figure}
\centerline{\includegraphics[width=0.8\textwidth]{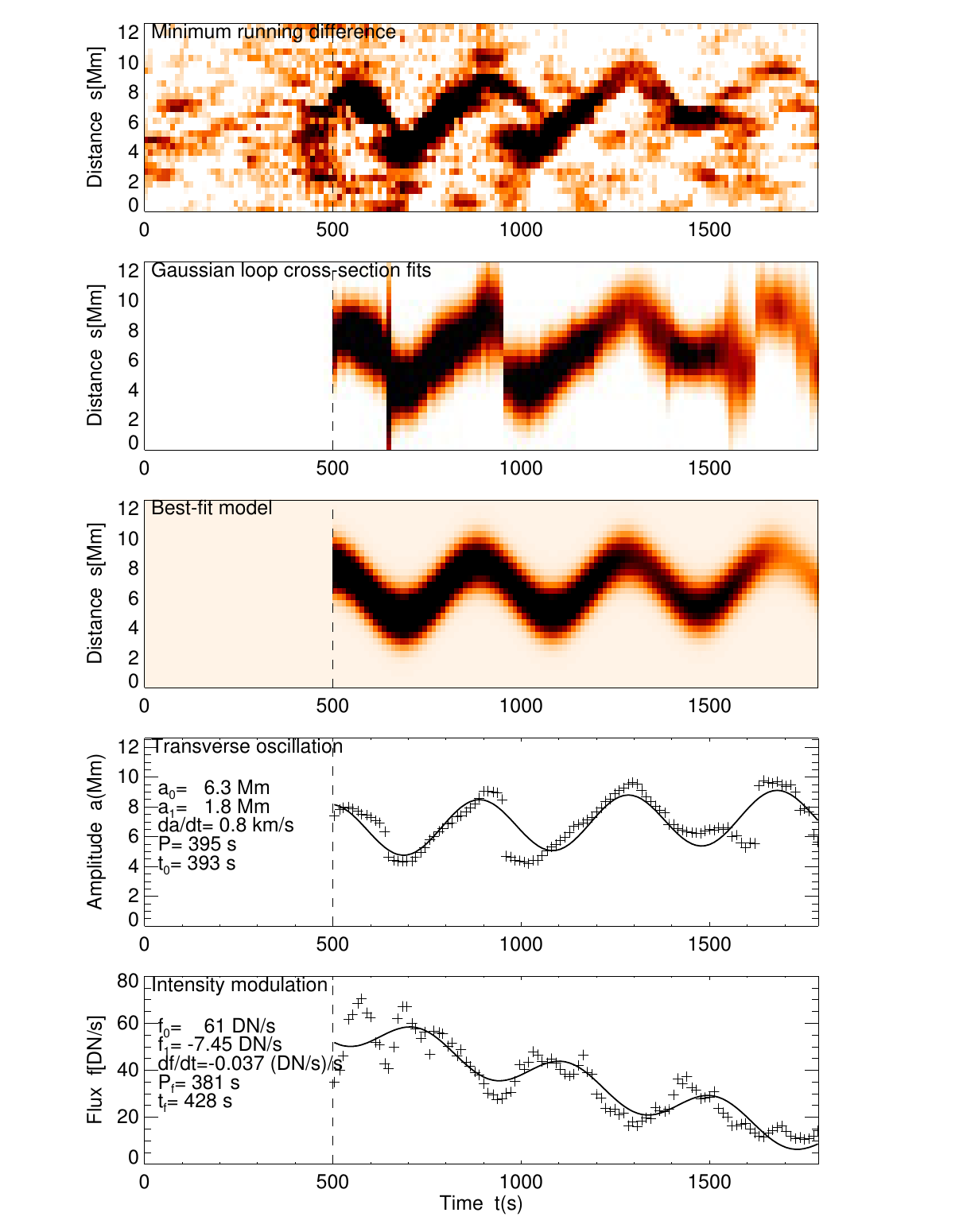}}
\caption{The analysis of the oscillation event of 
2010 October 16 is shown, with 
AIA/SDO running-difference time-slice plot (first panel),
Gaussian loop cross-section fits (second panel), best-fit
single loop oscillation profile $a(t)$ (third panel), 
transverse loop oscillation amplitude $a(t)$ (fourth panel), 
and EUV flux intensity modulation (fifth panel). Note that
the loop oscillation amplitude $a(t)$ is anti-correlated to the
intensity modulation flux $f(t)$ (Aschwanden and Schrijver 2011).}
\end{figure}
\clearpage

\begin{figure}
\centerline{\includegraphics[width=0.8\textwidth]{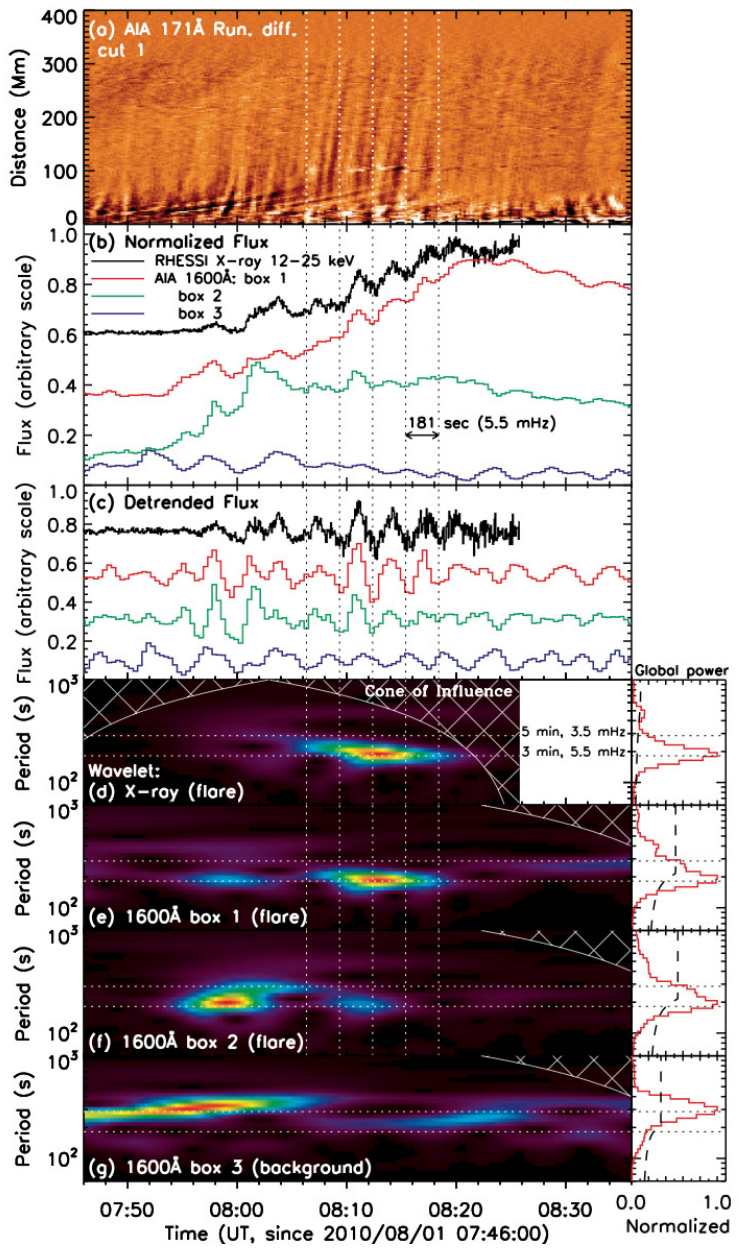}}
\caption{Quasi-periodic flare pulsations during the 
2010 August 1 C3.2 flare/CME event: (a) Time-slice plot
of AIA 171 \ang\ data. (b) RHESSI 12-25 keV X-ray flux
and AIA 1600 \ang\ fluxes. (c) Detrended fluxes obtained
by subtracting smoothed fluxes. (d-g) Wavelet power spectra.
Note the fast drift rates with phase speeds of 
$v \approx 2000$ km s$^{-1}$, which are interpreted as
fast-mode MHD waves that play an important role in the 
flare evolution (Liu et al.~2011).}
\end{figure}
\clearpage

\begin{figure}
\centerline{\includegraphics[width=0.8\textwidth]{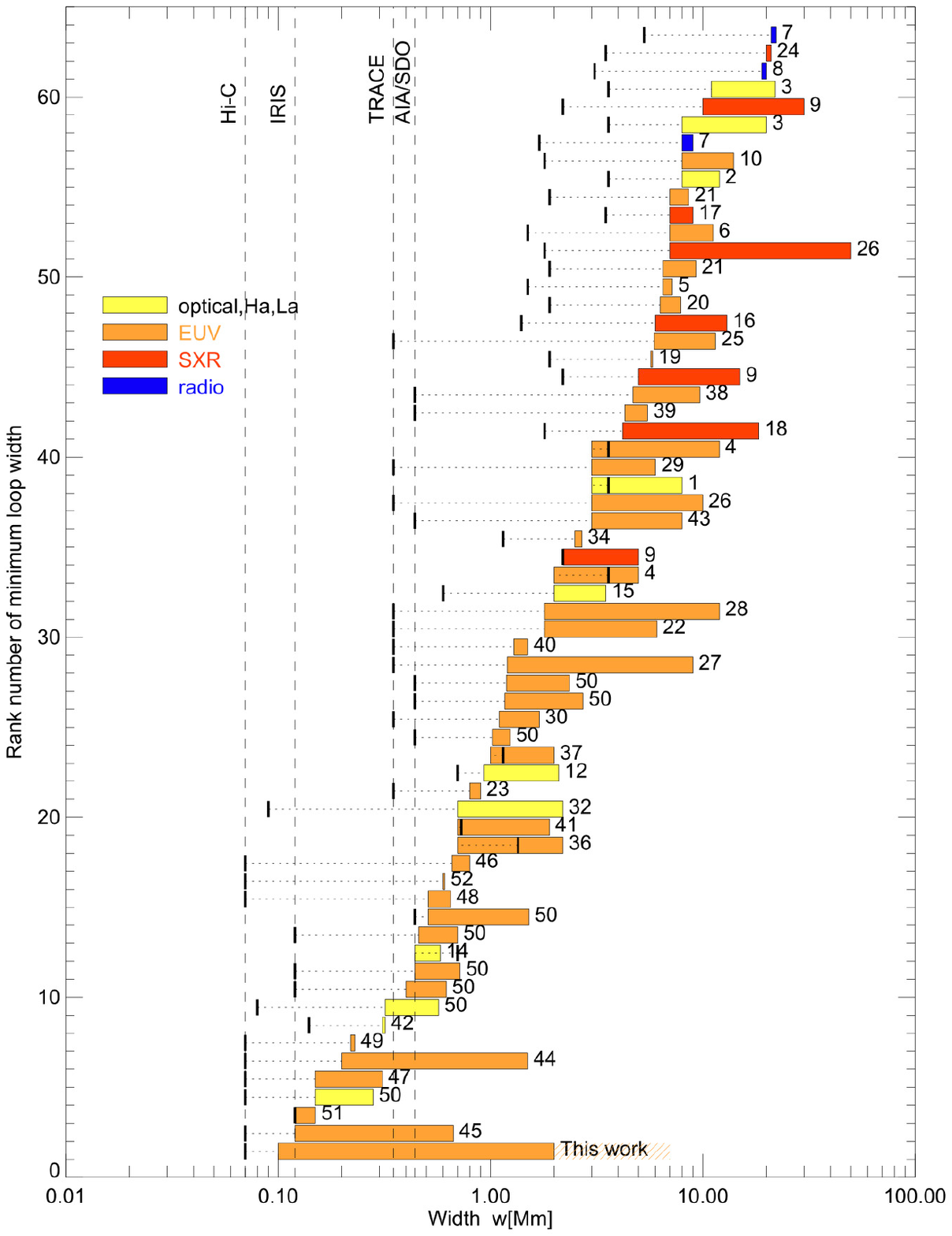}}
\caption{Synopsis of reported widths of coronal loops with
different wavelengths and instruments. Data are gathered
from 46 different publications during 1963-2017. The loop
withs $w$ are typically a factor of $\approx 2$ larger than
the used instrument resolutions (indicated with vertical bars).
The numbered references are listed in the original publication
of Aschwanden and Peter (2017).}
\end{figure}
\clearpage

\begin{figure}
\centerline{\includegraphics[width=0.8\textwidth]{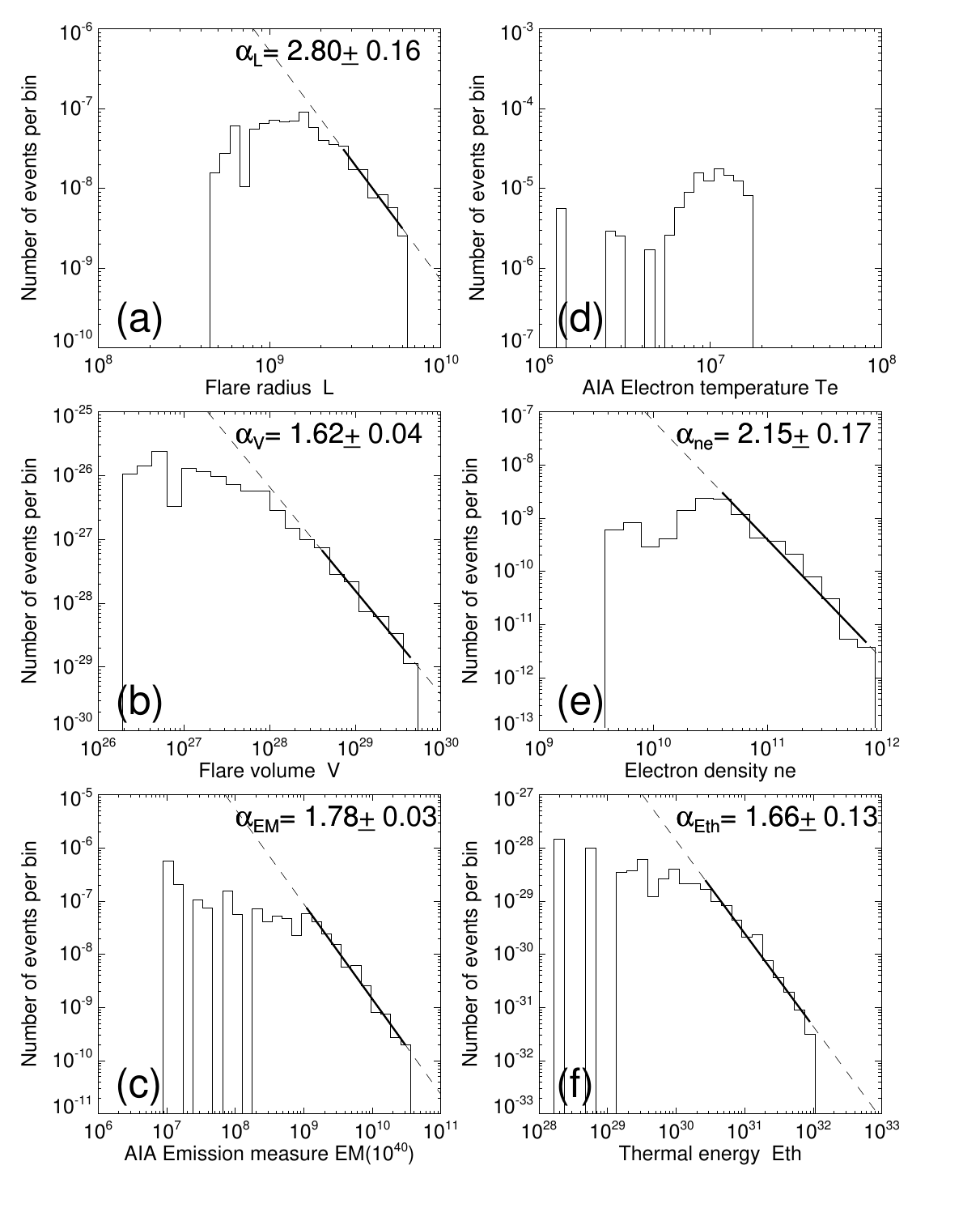}}
\caption{Size distributions of the flare length scale (a),
flare volume (b), emission measure (c), electron temperature (d),
electron density (e), and thermal energy (f) obtained from
155 M- and X-class flare events observed with AIA/SDO,
(Aschwanden and Shimizu 2013).}
\end{figure}
\clearpage

\begin{figure}
\centerline{\includegraphics[width=0.8\textwidth]{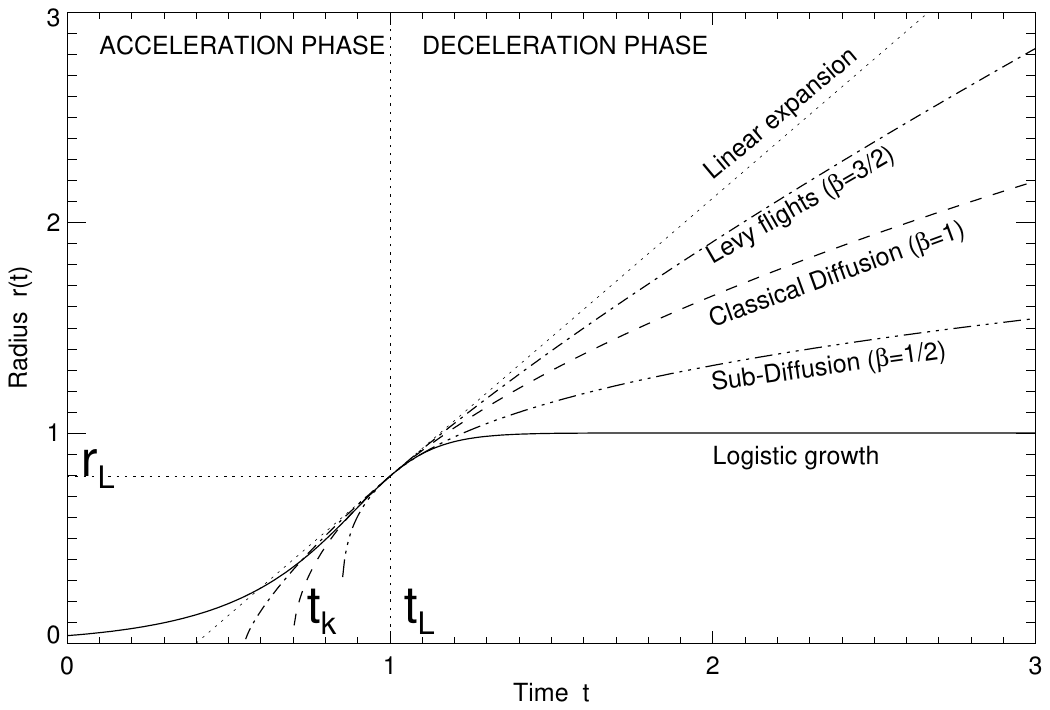}}
\caption{Comparison of spatio-temporal evolution models:
Logistic growth with parameters $t_L=1.0, r_\infty=1.0, \tau_G=0.1$,
sub-diffusion ($\beta=1/2$), classical diffusion ($\beta=1$),
L\'evy flights or super-diffusion ($\beta=3/2$), and linear expansion 
($r \propto t$).  All three curves intersect at $t=t_L$
and have the same speed $v=(dr/dt)$ at the intersection point at
time $t=t_L$, (Aschwanden et al.~2012b).}  
\end{figure}
\clearpage

\begin{figure}
\centerline{\includegraphics[width=0.8\textwidth]{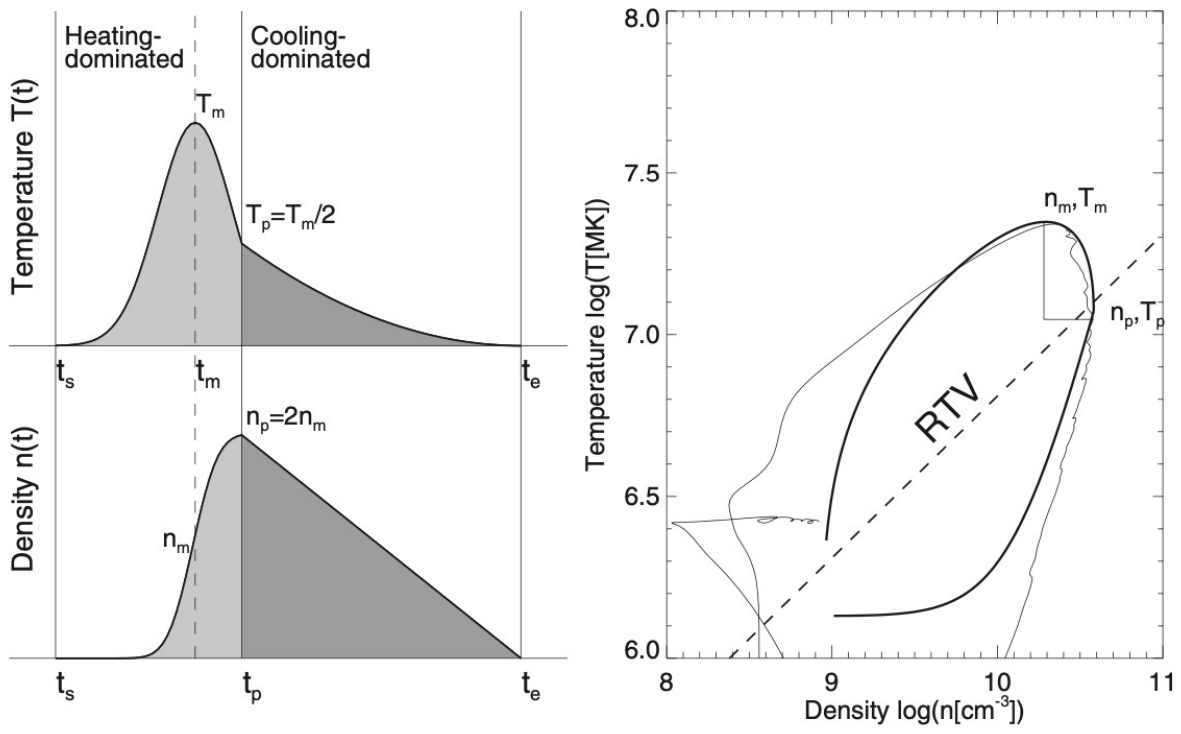}}
\caption{Hydrodynamic time evolution of the electron temperature 
$T(t)$ and density $n_e(t)$ of a simulation of an impulsively 
heated flare loop, shown as time profiles (left panel) and as an 
evolutionary phase diagram $T_e(n_e)$ (right panel). The 
evolution of the hydrodynamic simulation is shown
as exact numerical solution (thin line in right panel) and as an 
analytical approximation (thick lines in both panels), along with 
the prediction $T_e \propto n_e$ of the RTV scaling law for 
uniform steady heating (dashed line in right panel),
(Aschwanden et al.~2012b).}
\end{figure}
\clearpage

\begin{figure}          
\centerline{\includegraphics[width=1.0\textwidth]{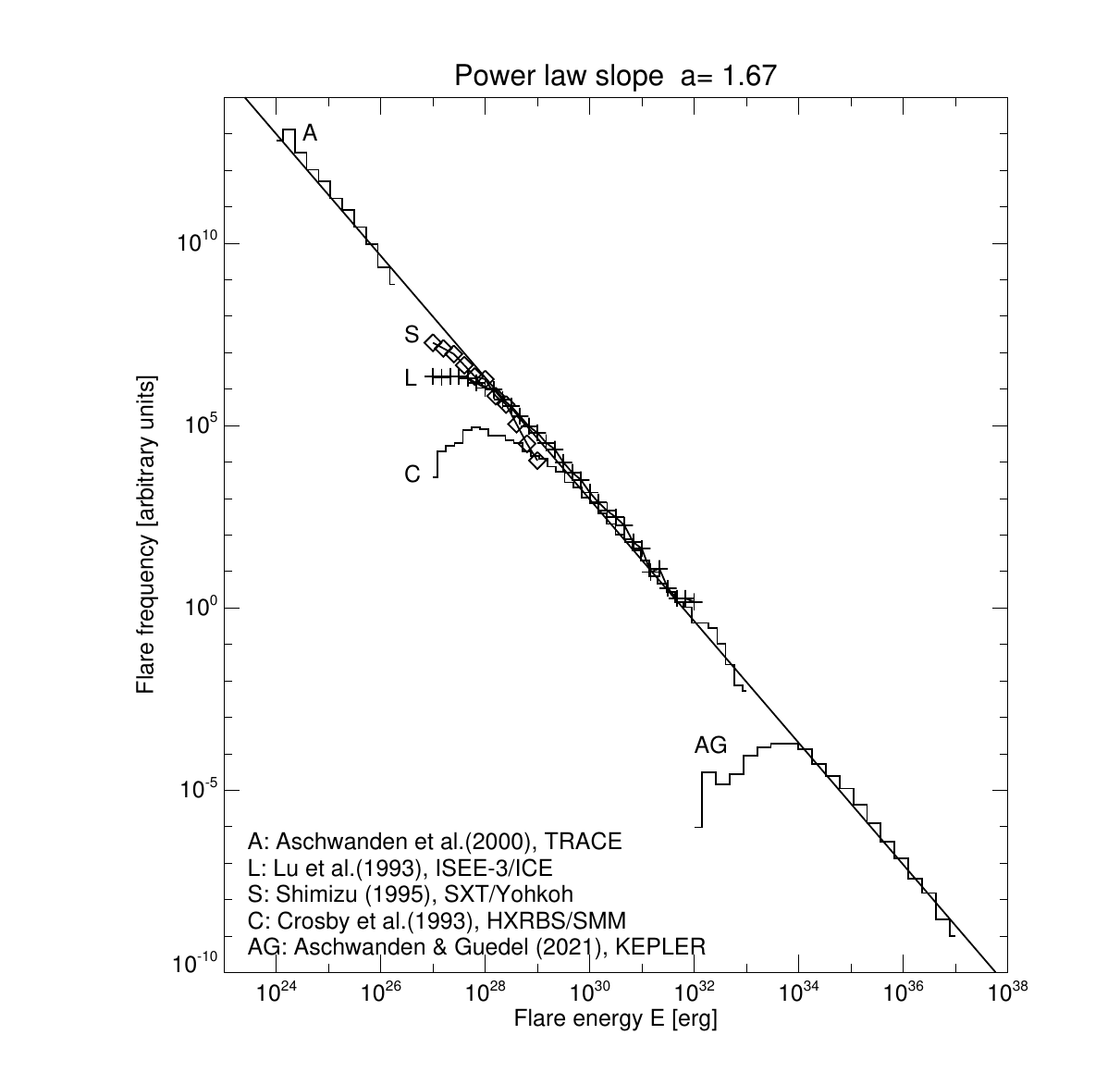}}
\caption{Energy size distributions synthesized from solar
observations with HXRBS/SMM, ISEE-3/ICE, SXT/Yohkoh, TRACE
and from stellar observaions with KEPLER. The theoretical
predictions of the power law slope for the flare energy
distribution is $\alpha_E = 1.67$ in the FD-SOC model.
The flare frequencies are aligned to the predicted power
law size distributions (Aschwanden and Schrijver 2025).} 
\end{figure}
\clearpage

\begin{figure}
\centerline{\includegraphics[width=0.8\textwidth]{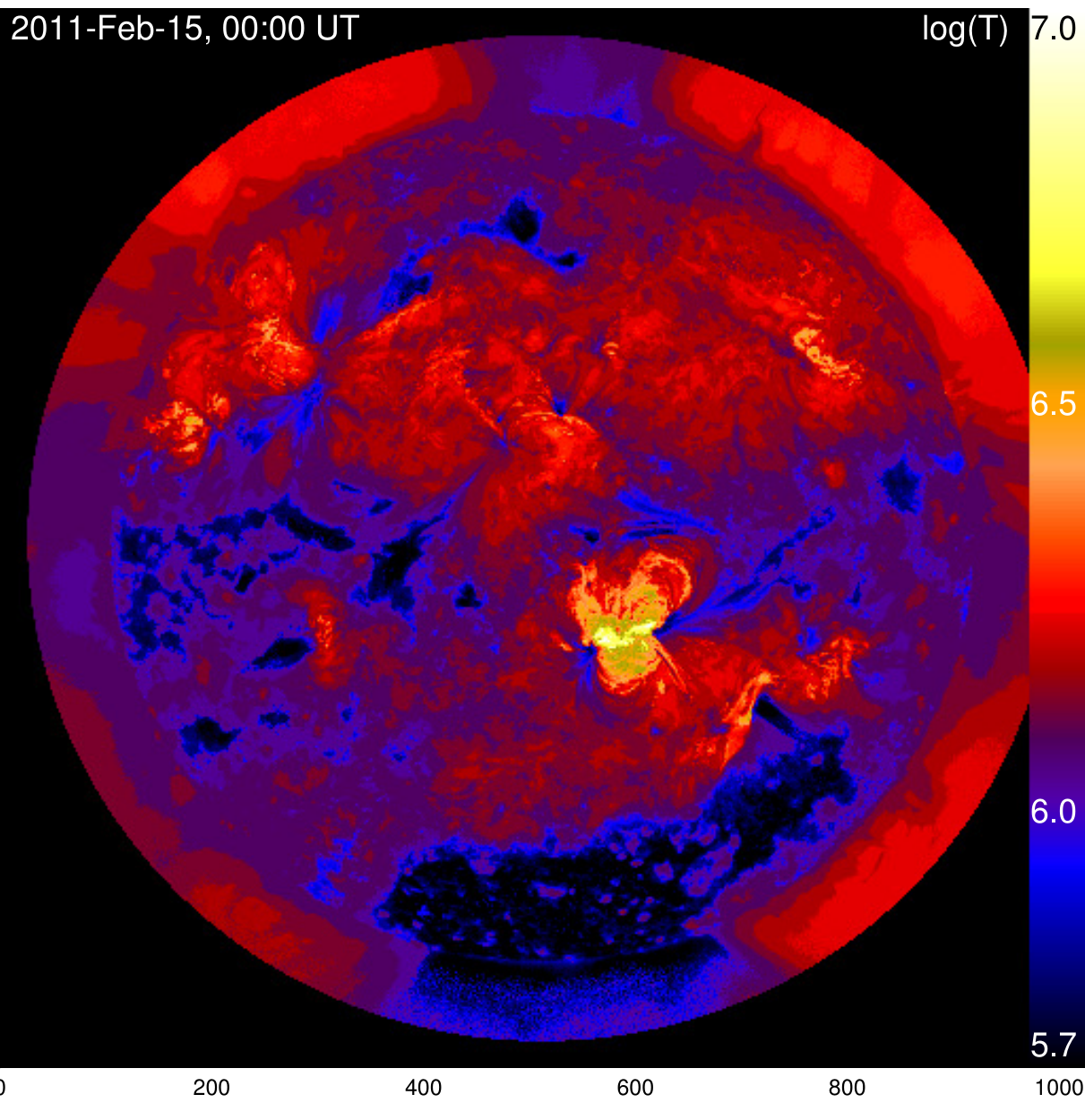}}
\caption{A temperature map calculated from the six AIA coronal filters
recorded on 15 February 2011, at 00:00 UT. The temperature range is
indicated in the vertical color bar on the right side,  
$\log(T)$ = 5.7\,--\,7.0. The spatial resolution of the temperature map
is $2.4\arcsec$ and each temperature value is calculated for an
averaged macropixel with an area of $4\times 4$ pixels
(Aschwanden et al.~2013b).}
\end{figure}
\clearpage

\begin{figure}
\centerline{\includegraphics[width=0.8\textwidth]{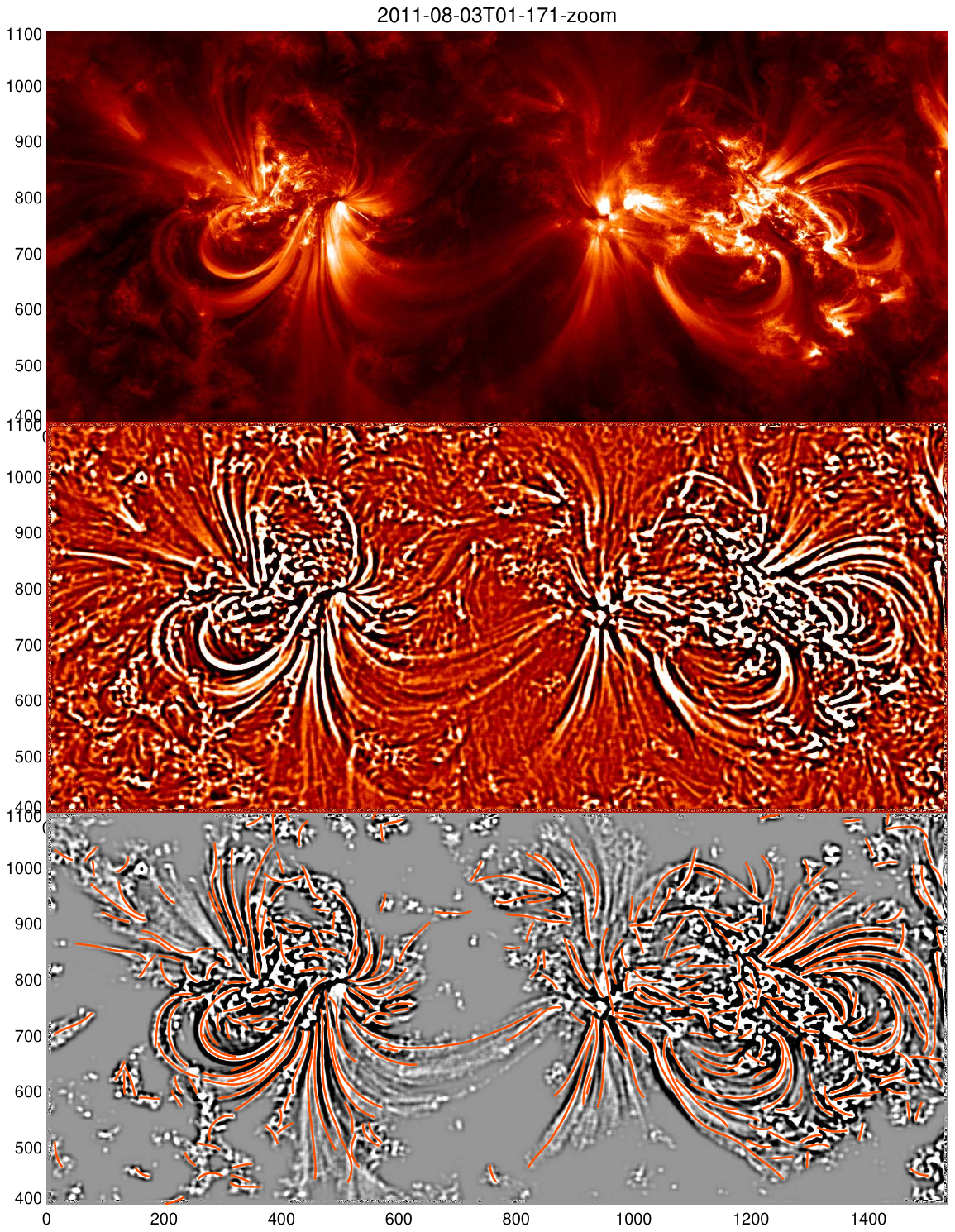}}
\caption{Bandpass-filtered image of an active region complex observed
with AIA/SDO on 2011 Aug 3, 01 UT, 171 \ang\ , shown as intensity image
(top panel), as bandpass-filtered version with $n_{sm1}=9$ and
$n_{sm2}=11$ (middle panel), and overlaid with automatically traced
loop structures (bottom panel), where the low-intensity values below
the median of $f=75$ DN are blocked out (grey areas),
(Aschwanden, DePontieu, and Katrukha 2013b).}
\end{figure}
\clearpage

\begin{figure}
\centerline{\includegraphics[width=0.8\textwidth]{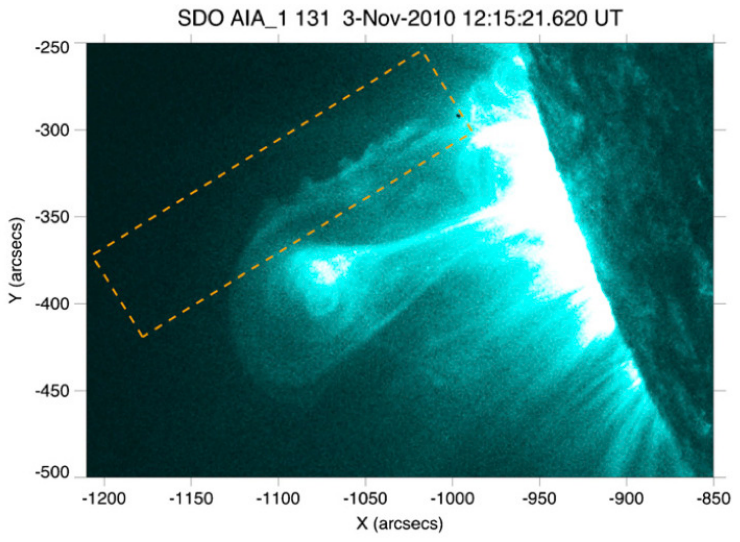}}
\caption{Fast coronal mass ejecta erupting from the Sun, with 
{\sl Kelvin-Helmholtz instability (KHI)}
waves detected on its northern flank. The SDO/AIA image 
is the 131 \ang\ channel. With increasing (brighter) intensity 
levels, it shows the ejecta canopy and within it, a brighter 
core above a thinner reconnecting current sheet. Note the
vortices at the boundary between the eruptive flare region 
and the ambient solar wind (Foullon et al~2011).}
\end{figure}
\clearpage

\begin{figure}
\centerline{\includegraphics[width=0.8\textwidth]{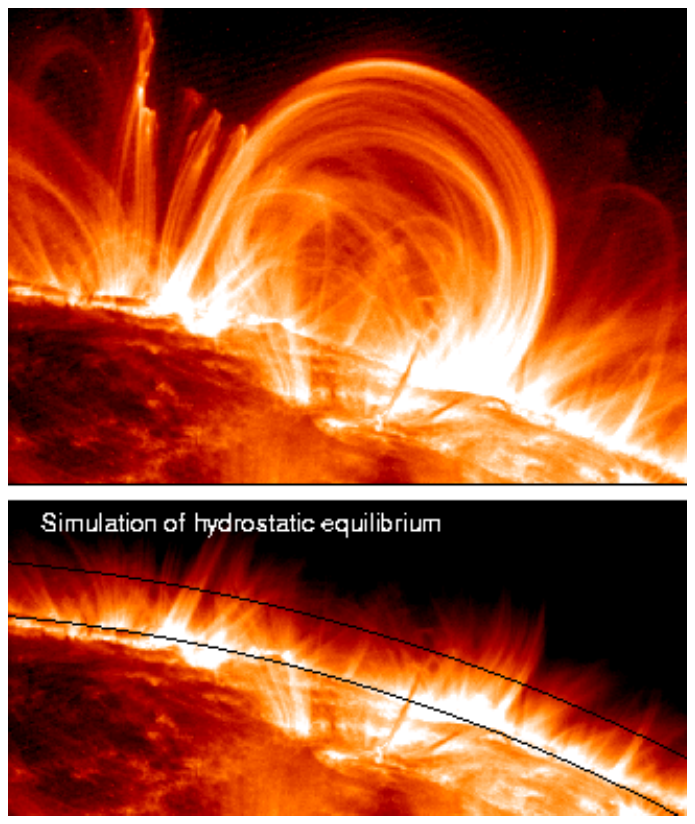}}
\caption{An active region with many loops that have an extended
scale heigth of $\lambda_p/\lambda_T < 3-4$ (top) has been scaled
to the hydrodynamic thermal scale height of $T=1$ MK (bottom).
The pressure scale height of the 1 MK plasma is $\lambda_T=
47,000$ km, but the observed flux is proportional to the
emission measure $(F \mapsto EM \mapsto n_e^2)$, which has
the half pressure scale height $\lambda/2=23,000$ km,
(Aschwanden 2004).}
\end{figure}
\clearpage

\begin{figure}
\centerline{\includegraphics[width=0.8\textwidth]{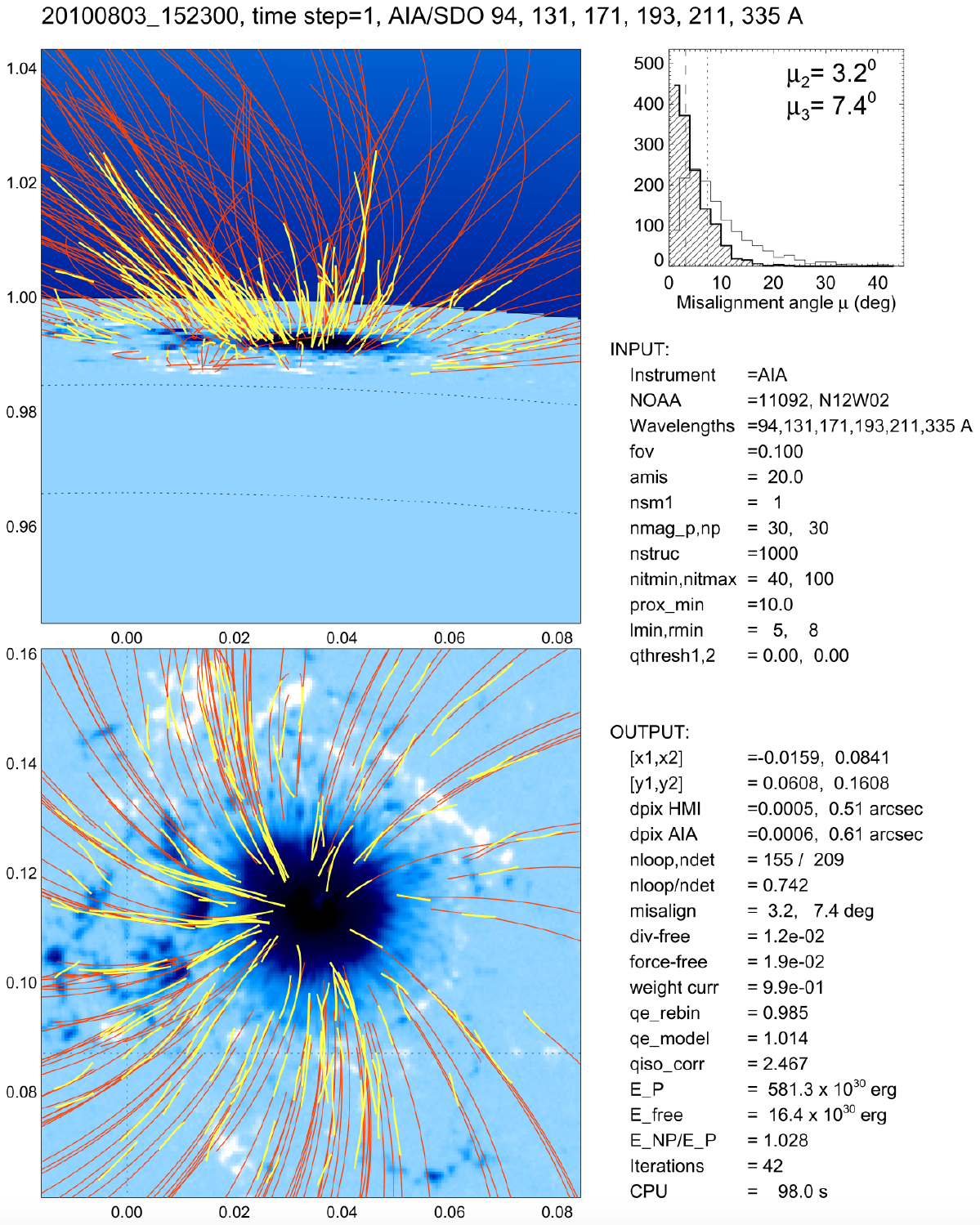}}
\caption{The automated curvi-linear feature tracing in the 
AIA images (2010 August 03, 15:23 UT) in 6 AIA wavelengths 
of 94, 131, 171, 193, 211, 335 Å (yellow curves) are shown, 
overlaid on the best-fit solutions of the magnetic field 
model using the VCA-NLFFF code (red curves), and the observed 
HMI magnetogram (blue background image), from the line-of-sight 
view in the (x, y)-plane (bottom panel) and the orthogonal 
projection in the (x, z)-plane (top panel). A histogram of the 2D
and 3D misalignment angles and various input and output 
parameters are shown in the top right-hand panel
(Aschwanden et al.~2016b).}
\end{figure}
\clearpage

\begin{figure} 
\centerline{\includegraphics[width=0.8\textwidth]{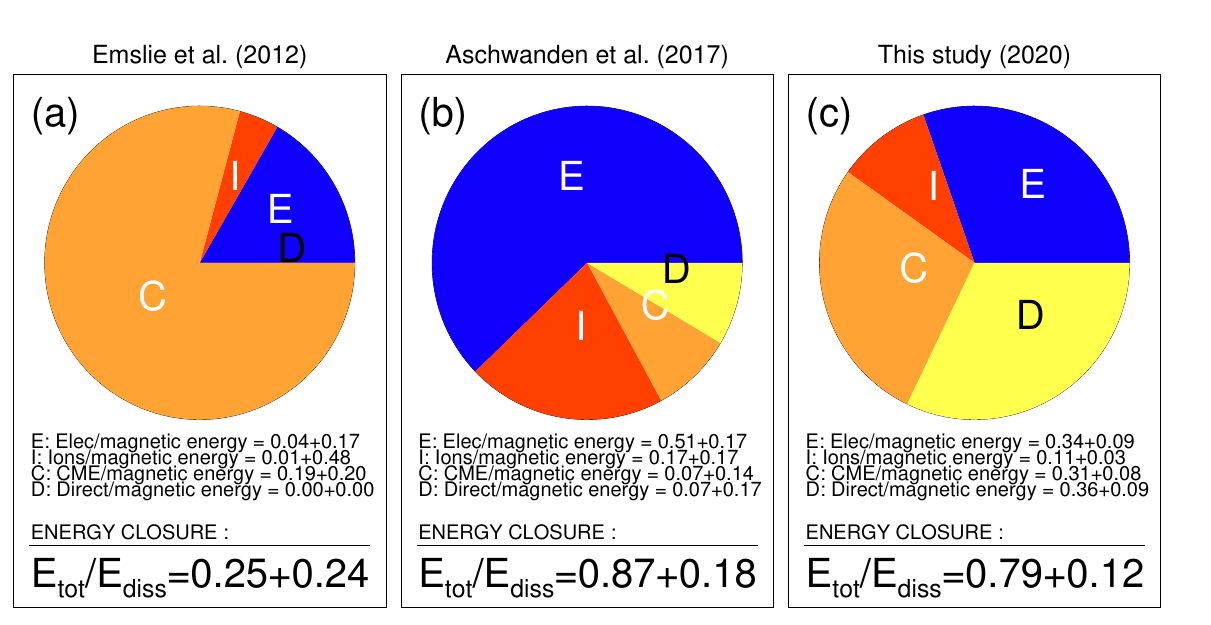}}
\caption{Pie chart diagrams the energy ratios of electrons
(blue), ions (red), CMEs (orange), and direct heating (yellow)
as a fraction of the total dissipated magnetic energy,
for the data sets of 
Emslie et al.~(2012) (left panel), 
Aschwanden et al.~(2017) (middle panel), and 
Aschwanden et al.~(2020a) (right panel).}
\end{figure}
\clearpage

\end{document}